\documentclass[fleqn,usenatbib]{mnras}

\usepackage{newtxtext,newtxmath}

\usepackage[T1]{fontenc}
\usepackage{ae,aecompl}

\usepackage{graphicx}	
\usepackage{amsmath}	
\usepackage{amssymb}	

\usepackage{amstext}
\usepackage[all]{hypcap}
\usepackage{afterpage}
\usepackage{gensymb}
\usepackage{color}

\usepackage[british]{babel}
\usepackage{multicol}
\usepackage{dblfloatfix}
\usepackage{enumitem}
\usepackage{soul}
\usepackage[flushleft]{threeparttable}
\usepackage{longtable}
\usepackage[dvipsnames]{xcolor}

\title[MWC 560]{Regulation of accretion by its outflow in a symbiotic star:\\the 2016 outflow fast state of MWC 560}

\author[A. B. Lucy et al.]{Adrian B. Lucy,$^{1}$\thanks{LSSTC Data Science Fellow}\thanks{E-mail: lucy@astro.columbia.edu (ABL)}
J. L. Sokoloski,$^{1,2}$
U. Munari,$^{3}$
Nirupam Roy,$^{4}$
N. Paul M. Kuin,$^{5}$
\newauthor
Michael P. Rupen,$^{6,7}$
Christian Knigge,$^{8}$
M. J. Darnley,$^{9}$
G. J. M. Luna,$^{10,11}$
\newauthor
P\'eter Somogyi,$^{12}$
P. Valisa,$^{13}$
A. Milani,$^{13}$
U. Sollecchia,$^{13}$
and Jennifer H. S. Weston$^{14}$
\\
$^{1}$Columbia University, Dept. of Astronomy, 550 West 120th Street, New York, NY 10027, U.S.A.\\
$^{2}$Large Synoptic Survey Telescope Corporation, 933 North Cherry Ave, Tucson, AZ 85721, USA\\
$^{3}$INAF Astronomical Observatory of Padova, 36012 Asiago (VI), Italy\\
$^{4}$Department of Physics, Indian Institute of Science, Bangalore 560012, India\\
$^{5}$Mullard Space Science Laboratory, University College London, Holmbury St. Mary, Dorking, Surrey RH5 6NT, UK\\
$^{6}$National Radio Astronomy Observatory, Socorro, New Mexico 87801, USA\\
$^{7}$National Research Council, Herzberg Astronomy and Astrophysics, 717 White Lake Road, PO Box 248, Penticton, BC V2A 6J9, Canada\\
$^{8}$University of Southampton, School of Physics \& Astronomy, Highfield, Southampton, SO17 1BJ, U.K.\\
$^{9}$Astrophysics Research Institute, Liverpool John Moores University, Liverpool, L3 5RF, UK\\
$^{10}$CONICET-Universidad de Buenos Aires, Instituto de Astronom\'{\i}a y F\'{\i}sica del Espacio (IAFE),\\ ~~Av. Inte. G\"{u}iraldes 2620, C1428ZAA, Buenos Aires, Argentina\\
$^{11}$Universidad de Buenos Aires, Facultad de Ciencias Exactas y Naturales, Buenos Aires, Argentina\\
$^{12}$Astronomical Ring for Access to Spectroscopy\\
$^{13}$ANS Collaboration, c/o Astronomical Observatory, 36012 Asiago (VI), Italy\\
$^{14}$AAAS S\&T Policy Fellow, National Science Foundation, USA
}

\date{Submitted May 6, 2019}

\pubyear{2019}

\begin{document}
\label{firstpage}
\pagerange{\pageref{firstpage}--\pageref{lastpage}}
\maketitle

\begin{abstract}
How are accretion discs affected by their outflows? To address this question for white dwarfs accreting from cool giants, we performed optical, radio, X-ray, and ultraviolet observations of the outflow-driving symbiotic star MWC~560 ($\equiv$V694~Mon) during its 2016 optical high state. We tracked multi-wavelength changes that signalled an abrupt increase in outflow power at the initiation of a months-long outflow fast state, just as the optical flux peaked: {\it(1)} an abrupt doubling of Balmer absorption velocities; {\it(2)} the onset of a $20$~$\mu$Jy/month increase in radio flux; and {\it(3)} an order-of-magnitude increase in soft X-ray flux. Juxtaposing to prior X-ray observations and their coeval optical spectra, we infer that both high-velocity and low-velocity optical outflow components must be simultaneously present to yield a large soft X-ray flux, which may originate in shocks where these fast and slow absorbers collide. Our optical and ultraviolet spectra indicate that the broad absorption-line gas was fast, stable, and dense ($\gtrsim10^{6.5}$\,cm$^{-3}$) throughout the 2016 outflow fast state, steadily feeding a lower-density ($\lesssim10^{5.5}$\,cm$^{-3}$) region of radio-emitting gas. Persistent optical and ultraviolet flickering indicate that the accretion disc remained intact. The stability of these properties in 2016 contrasts to their instability during MWC~560's 1990 outburst, even though the disc reached a similar accretion rate. We propose that the self-regulatory effect of a steady fast outflow from the disc in 2016 prevented a catastrophic ejection of the inner disc. This behaviour in a symbiotic binary resembles disc/outflow relationships governing accretion state changes in X-ray binaries. \end{abstract}

\begin{keywords}
binaries: symbiotic -- 
stars: winds, outflows -- stars: jets -- accretion, accretion discs -- stars: individual (MWC 560) -- white dwarfs
\end{keywords}



\vspace*{2.0cm}

\section{Introduction} \label{intro}

White dwarf (WD) symbiotic stars, hereafter symbiotics, are interacting binaries in which a WD accretes from a cool giant. They have larger accretion discs than their counterparts with main-sequence-like donors (cataclysmic variables: CVs), and several have spatially resolved jets. They are candidate progenitors for single-degenerate supernovae Ia \citep[e.g.,][]{Cao2015,Munari1992} and possible ancestors of double-degenerate supernovae Ia, raising the stakes for understanding their accretion discs, WD masses, and evolution. Accretion disc outflows may be fundamental to that investigation.

MWC 560 ($\equiv$V694 Mon) is a remarkable symbiotic in which the accretion disc drives a powerful outflow, producing broad, blue-shifted, variable absorption lines that extend up to thousands of km s$^{-1}$ from atomic transitions in the infrared (IR), optical, near-ultraviolet (NUV), and far-ultraviolet (FUV): \ion{He}{i}, \ion{H}{i}, \ion{Al}{iii}, \ion{Mg}{ii}, \ion{Fe}{ii}, \ion{Cr}{ii}, \ion{Si}{ii},\ion{C}{ii}, \ion{Ca}{ii}, \ion{Mg}{i}, \ion{Na}{i}, \ion{O}{i}, \ion{C}{iv}, \ion{Si}{iv}, and \ion{N}{v} \citep{Bond1984,Tomov1990,Iijima2001,Michalitsianos1991,Tomov1992,Meier1996,Schmid2001,Goranskij2011,Lucy2018}. It may be the prototype of a population of broad absorption line symbiotics \citep{Lucy2018}, and a useful nanoscale-mass analogy to quasars both in terms of its emission lines \citep{Zamanov2002} and its absorption lines \citep{Lucy2018}. The distance to the system is probably about 2.5~kpc (Appendix~\ref{app1}). The red giant (RG) is mid-M and not a Mira; the infrared spectrum is S-type in that it is dominated by the RG rather than by dust \citep{Meier1996}. The WD mass is probably at least 0.9 M$_{\odot}$ (\citealt{Zamanov2011a}, \citealt{Stute2009}). The outflow may be a highly-collimated, baryon-loaded jet \citep{Schmid2001} or a polar wind with a wider opening angle \citep{Lucy2018}. \citet{Zamanov2011a} have previously proposed that switching between different mass outflow regimes in MWC 560 is related in some way to switching between different regimes of mass inflow in the inner accretion disc, but the nature of that relationship is undetermined.

Here we describe coordinated multi-wavelength observations of MWC 560 conducted during 2016, prompted by the January 2016 peak of a year-long rise in optical flux (which may have been predicted by the system's flux periodicities; \citealt{Leibowitz2015}). Our 2016 observations are supplemented by data collected throughout the preceding decade. In \autoref{observations}, we describe our observations and data reduction in the optical (\autoref{obsoptical}), radio (\autoref{obsradio}), X-ray (\autoref{obsxrays}), and NUV (\autoref{obsUV}) wavebands. We present our results in \autoref{results}, examining the optical absorption (\autoref{opticalabs}), the radio flux (\autoref{resultsradio}), the X-ray spectrum (\autoref{resultsxrays}), the NUV absorption (\autoref{uvspectrasec}), and the optical and NUV flickering (\autoref{flickering}). We cohere our results into a physical narrative in \autoref{discussion}, describing the state of the outflow (\autoref{4.1}) and of the accretion disc (\autoref{4.2}), then demonstrating a regulatory relationship between the disc and its outflow over the course of MWC 560's history (\autoref{4.3}). We summarize our conclusions in \autoref{conclusions}.

Throughout this paper, we use heliocentric velocities, and scale to a distance of 2.5 kpc and a WD mass of 0.9 M$_{\odot}$ unless otherwise noted.

\section{Observations and Data Reduction} \label{observations}

\subsection{Optical spectroscopy} \label{obsoptical}

We obtained 74 optical spectra of MWC 560\footnote{Including one spectrum presented previously in \citet{Munari2016}.}, listed in Appendix~\ref{app2} and delineated below.

We obtained 30 echelle spectra with two telescopes at Asiago: the ANS
Collaboration 0.61m operated in Varese by Schiaparelly Observatory,
and the 1.82m operated by the National Institute of Astrophysics
(INAF). The Varese 0.61m telescope used an Astrolight Instruments mark.III Multi Mode Spectrograph with a SBIG ST10XME CCD camera, covering 4225--8900 \AA\ in 27 orders with a spectral resolution of R=18000,
The Asiago 1.82m telescope used the
REOSC Echelle spectrograph with an Andor DW436-BV and
an E2V CCD42-40 AIMO back illuminated CCD, covering 3600--7300\AA\ in 32
orders with R=20000.

The wavelength calibration for each echelle spectrum was obtained by
exposing on a Thorium lamp before and after the science spectrum. A pruned
list of about 800 unblended Thorium lines evenly distributed along and among the
Echelle orders was fitted to the observed spectra, providing a wavelength
solution with an rms of 0.005 and 0.009 \AA\ ($\equiv$0.3 and 0.6 km s$^{-1}$)
for the Asiago and Varese spectrographs, respectively. 

We also obtained 16 spectra with R$\approx$1300 (3300--8050~\AA) at Asiago with the 1.22m telescope + B\&C spectrograph, operated in Asiago by the University of Padova, using an ANDOR iDus DU440A with a back-illuminated E2V 42-10 sensor. 

At all Asiago telescopes, the long slit, with a 2 arcsec slit width, was aligned with the parallactic
angle. Spectrophotometric standards at similar airmass were observed immediately
before and after MWC 560. The spectra were similarly reduced within IRAF,
carefully involving all steps connected with bias correction, darks and
flats, long-slit sky subtraction, and wavelength and flux calibration. 

We obtained 10 spectra with R$\approx$1300 (3750-7350 \AA) using a home-built slit-spectrograph mounted on an 8-in C8 Schmidt-Cassegrain telescope located in L'Aquila, Italy. The slit
 width was set to 3 arcsec and a fixed East-West orientation. The wavelength
 solution was obtain by fitting to an Ar-Ne comparison lamp, and observation of
 spectrophotometric standard stars was used to flux the spectra. Due to the consistency of the data reduction process for these spectra and the close match to other spectra when they were taken at similar times, we will hereafter group these data with the 1.22m telescope +
B\&C spectrograph spectra, which have the same resolving power.

We obtained 10 H$\alpha$ spectra of varying resolutions with a Shelyak LHires III spectrograph, using 150--2400~gr~mm$^{-1}$ gratings and an ATIK 414 EXm camera, on 25 and 30 cm Newtonian telescopes located in Tata, Hungary. Slit widths varied from 15 to 35 microns. These spectra were reduced in ISIS following the standard procedure. The response calibrator star was fit with a 3rd order polynomial for spectra with R$\sim$3000 and above, and a manually selected spline for lower-resolution spectra only. The response calibration does not exhibit any features that would give rise to results discussed in this paper.

We obtained 8 spectra with the 2m fully robotic Liverpool Telescope (LT; \citealt{Steele2004}) using the FRODOSpec instrument \citep{Barnsley2012}. FRODOSpec was operated in its ``low resolution'' mode, obtaining spectra covering 3900--5700\,\AA\ and 5800--9400\,\AA\ at a resolution of $R\sim2500$. Each spectra epoch consisted of $3\times60$\,s exposures. Wavelength calibration was performed by comparison to a Xe arc lamp and the data were reduced using the pipeline described in \citet{Barnsley2012}.

\subsection{Radio} \label{obsradio}

In 2016, we observed for a total of 18 hours on the VLA through Projects 16A-448 and 16A-490, divided between 5 epochs 
in the S band (3.1 GHz), 7 epochs in the X band (9.8 GHz) and 4 epochs in the Ka band (33.1 GHz). Additionally, on 2014 October 2, we observed MWC 560 in the X band using about 2 hours of VLA Project 14B-394. More observational details are listed in \autoref{tableradio}.

\begin{table*}
\begin{threeparttable}
\begin{tabular}{lccccccc}
\hline
Date \tnote{a} & Config.\tnote{b} & Band & $\bar{\nu}$ (GHz) \tnote{c} & On-source time (min) & Flux density ($\mu$Jy) & $\sigma_{rand}$ ($\mu$Jy) \tnote{d} & $\sigma_{tot}$ ($\mu$Jy) \tnote{e}\\
\hline
2014 Oct 02 & DnC & X & 9.83 & 88 & 37 & 3 & 4 \\
2016 Apr 04 & C & X & 9.84 & 68 & 85 & 4 & 6 \\
2016 May 01 & CnB & X & 9.84 & 66 & 105\tnote{f} & 19\tnote{f} & 20 \\
2016 May 24 & B & X & 9.86 & 67 & 136 & 4 & 8 \\
2016 Jul 29 & B & X & 9.82 & 40 & 175 & 5 & 10 \\
2016 Oct 17 & A & X & 9.85 & 40 & 144 & 4 & 9 \\
2016 Nov 09 & A & X & 9.85 & 45 & 138 & 4 & 8 \\
2017 Jan 18 & A & X & 9.80 & 42 & 120 & 4 & 7 \\
\\
2016 May 24 & B & S & 3.10 & 67 & 128 & 5 & 8 \\
2016 Jul 29 & B & S & 3.08 & 38 & 163 & 7 & 11 \\
2016 Oct 17 & A & S & 3.07 & 38 & 162 & 7 & 11 \\
2016 Nov 09 & A & S & 3.08 & 37 & 135 & 7 & 10 \\
2017 Jan 18 & A & S & 3.11 & 37 & 112 & 6 & 8 \\
\\
2016 Jul 29 & B & Ka & 33.07 & 22 & 183 & 16 & 24 \\
2016 Oct 12 & A & Ka & 33.07 & 21 & 153 & 10 & 19 \\
2016 Nov 09 & A & Ka & 33.09 & 26 & 163 & 10 & 19 \\
2017 Jan 18 & A & Ka & 32.96 & 24 & 128 & 11 & 17 \\
\hline
\end{tabular}
\caption{VLA observations and results. All detections were unresolved point sources. \label{tableradio}}
\begin{tablenotes}
\item[a]\footnotesize UT date at block start. All S and Ka band observations were conducted within 3 hours of an X band observation and of each other, except the 2016 Oct 12 Ka band observation, which was obtained 5 days before the other bands.
\item[b]\footnotesize VLA configuration.
\item[c]\footnotesize Central frequency after flagging.
\item[d]\footnotesize Statistical uncertainty output by AIPS.
\item[e]\footnotesize Total uncertainty: statistical uncertainty propagated in quadrature with an estimate of the systematic uncertainty as described in the text.
\item[f]\footnotesize Flux density and uncertainty for 2016 May 1 were inferred as described in the text due to a weather-related problem.
\end{tablenotes}
\end{threeparttable}
\end{table*}

We observed 3C147 for flux/gain calibration in all bands, QSO J0730-116 for phase calibration in S and X, J0724-0715 for phase calibration in Ka, and J0319+4130 for polarization leakage calibration in the later epochs. Source/phase-calibrator switching was performed at the recommended rates for each band and antennae configuration; pointing observations and slews to the satellite free zone were used when appropriate. 

We used the default correlator setups for broadband continuum observations in all bands, including 8-bit samplers in S-band and 3-bit samplers in X and Ka bands, full polarization products, and effective bandpass sizes (after data reduction) of 2.05 (S), 4.1 (X), and about 8.1 (Ka) GHz.

We reduced the data following the standard analysis procedure 
using the Common Astronomy Software Applications (CASA) and the Astronomical 
Image Processing System (AIPS) software packages. The initial flagging and 
calibration used the VLA Scripted Calibration Pipeline v1.3.8 
in CASA v4.6.0. For the mixed set-up observations, after initial 
flagging, the data for different bands were selected and separated into multiple 
files by running the CASA task {\it split}, before running the pipeline for each 
band separately. The pipeline output was inspected critically, and, in some 
cases, further manual flagging (and then pipeline re-calibration) was also performed. After flagging and calibration, the visibility data for the 
target field were converted using the CASA task {\it exportuvfits} to standard 
FITS format for imaging and self-calibration, which was then performed in AIPS (version 
31DEC16) using the AIPS tasks IMAGR and CALIB. CLEANing boxes (including strong sources in outlier fields) were used for faster convergence in imaging and deconvolution. The ``phase-only'' self-calibration was done using the CLEAN 
components iteratively. The final images thus obtained were used to 
estimate the flux densities and associated statistical uncertainties using JMFIT in AIPS. 

As the target source in all 
our observations was unresolved, in JMFIT we set DOWIDTH = -1 (which 
adopts the restoring CLEAN beam shape for the source) to get robust estimates 
of the flux densities. For the S and X band 
data, we also checked the rough consistency of flux densities for other nearby sources 
in the field. The random uncertainty output by AIPS was propagated in quadrature with an estimated systematic uncertainty of $5\%$ (S and X bands) or $10\%$ (Ka band) of the measured flux, following \citet{Weston2016a}, to obtain an estimated total uncertainty $\sigma_{tot}$.

The X band observation on 2016
May 1 had some unidentified weather-related problem\footnote{The problem was likely a combination of bad weather, bad ionosphere, and low inclination. Gusty winds approached the recommended X band limit of 15 m s$^{-1}$, and towards the end of the observation 7 antennae started auto-stowing in sequence for up to 28 min due to high winds. API RMS phase reached at least 26.4$\degree$, close to the 30$\degree$ limit. The target elevation ranged from 25 to 11$\degree$, close to the recommended limit.},
and we were 
unable to estimate the flux density of the target robustly for this epoch. 
However, comparing the flux density of three other sources in the field with that 
of the adjacent epochs, finding them to be too dim by a factor $\approx$3.6, and scaling both the target flux and the observed RMS noise value by 3.6, we obtained an estimate with a large uncertainty. Reassuringly, the estimated value is intermediate  between the flux densities observed on April 4 and May 24. 

\subsection{X-rays} \label{obsxrays}

Chandra observed the system in Cycle 17 for 24.76 ks on 2016 March 8.285 UT and 24.76 ks on 2016 March 9.079 UT, using chip S3 on the ACIS-S array in VFAINT mode without a grating. 

We reprocessed the data following standard CIAO (v4.8, CALDB v4.7.2) procedures. Chandra detected MWC 560 as a point source always within 1 pixel (0.5 arcsec) of the expected coordinates on each dimension. We used {\it dmstat} to obtain the centroid source position (to accommodate uncertainty in the aspect solution and the expected position; for both observations, the calculated position was 7$^{h}$25$^{m}$51$^{s}$.322,-7$\degree$44'07''.99),  then extracted PSF-corrected spectra with {\it specextract} (with weight=no correctpsf=yes) using a constant circular source aperture radius of 1.93 arcsec (the 90\% enclosed count fraction radius at 6.4 keV for the Chandra PSF, which is smaller at lower energies) and a circular background annulus with inner and outer radii of 18 and 30 arcsec. As a check, we also extracted the Chandra radius with a 2.5 arcsec source aperture radius for comparison. The exposure time of good data was 49.3 ks. We extracted light curves from Chandra for the same source and background regions using {\it dmextract} with 1, 5, and 12.5 ks bins.

We observed MWC 560 with the Neil Gehrels Swift Observatory for 48 ks, and reduced data from the Swift X-ray Telescope (XRT; \citealt{Burrows2005}) using the online product-building tool\footnote{\url{www.swift.ac.uk/user_objects/}}. The source was only marginally detected at less than 3$\sigma$, so we did not attempt a centroid. We used the default grade range, and downloaded spectra corresponding to the average of all Swift data (2016 March 2 -- June 1) and of the last month (2016 May 5 -- June 1). 

For comparison purposes, we obtained from the XMM-Newton archive an unpublished X-ray spectrum observed on 2013 April 12 (PI: Stute), reduced following standard procedures in the Science Analysis Software (SAS).

\subsection{Ultraviolet} \label{obsUV}

\subsubsection{UV spectroscopy}

We obtained 17 low-resolution (R$\sim150$) UV Grism spectra (1700--2900\,\AA) on Swift UVOT \citep{Roming2005} at roughly regular intervals between 2016 March 2 -- June 1, using a total of 15 ks on Swift; the dates are listed in Appendix~\ref{app2}. 

These data were reduced using the Swift UVOT Grism UVOTPY package \citep[version 2.1.3, Kuin 2014, an implementation of the calibration from][]{Kuin2015}. The spectra were extracted with default parameters and the locations of zeroth order contamination were determined by inspecting the images. These contaminated regions are approximations; features (especially those near contaminated regions) that did not recur between spectra obtained at different roll angles may be spurious. The contaminated regions and a few whole-spectrum quality flags are tabulated in Appendix~\ref{app2}. The wavelength scale was shifted to match the IUE spectrum LWP19113 from 1990 November 2 \citep{Michalitsianos1991}, which we obtained from the IUE archive. We focused on the 2100-2900\AA\, range for comparison to older observations and to avoid auto-flagged short-wavelength data and long-wavelength second-order contamination.

We omit from our analysis three spectra noted in Appendix~\ref{app2} as having nearby bright first order flux or as suffering from underestimated flux due to loss of lock, leaving 14 good spectra.

\subsubsection{UV photometry}

We observed MWC 560 with the UVOT UVM2 ($\lambda_{cen}$=2246\AA, FWHM=513\AA) photometric filter on Swift for 33.6 ks in event mode from 2016 Mar--Jun, using a 5x5 arcmin hardware window to minimize coincidence loss.

Reductions were performed in HEAsoft. The data were screened using a development version of {\it uvotscreen} (the stable version at the time, HEAsoft v6.21, could apply the orbit file incorrectly) using \texttt{aoexpr="ELV > 10. \&\& SAA == 0"} and allowing quality flags of only 0 and 256. The photometry was extracted following the method described in \citet{Oates2009}, and we checked the images at each time step. A development version of {\it uvotsource}, patched to eliminate data from debris-shadowed regions on the detector\footnote{\url{https://heasarc.gsfc.nasa.gov/docs/heasarc/caldb/swift/docs/uvot/uvotcaldb_sss_01.pdf}}, was employed using a circular aperture of usually 5 arcsec radius (with larger radii during periods of mediocre tracking). We then binned the extracted data into 60 second intervals, propagating the errors appropriately. The same procedures were performed on a reference star, TYC 5396-570-1.

Our method yielded 21.6 ks of good live exposure on MWC 560. As a check, we also performed a more standard reduction using a stricter attitude filtering expression\footnote{\texttt{aoexpr="ANG\_DIST < 100. \&\& ELV > 10. \&\& SAA == 0 \&\& SAFEHOLD==0 \&\& SAC\_ADERR<0.2 \&\& STLOCKFL==1 \&\& STAST\_LOSSFCN<1.0e-9 \&\& TIME\{1\} - TIME < 1.1 \&\& TIME - TIME\{-1\} < 1.1"}}, manual removal of additional intervals with loss of lock or mediocre tracking, and a strict 5 arcsec radius aperture. An additional few thousand seconds of observation were lost using this check method, but all long-term and short-term variability trends described in this paper were still observed.

\begin{figure*}
\includegraphics[width=7in]{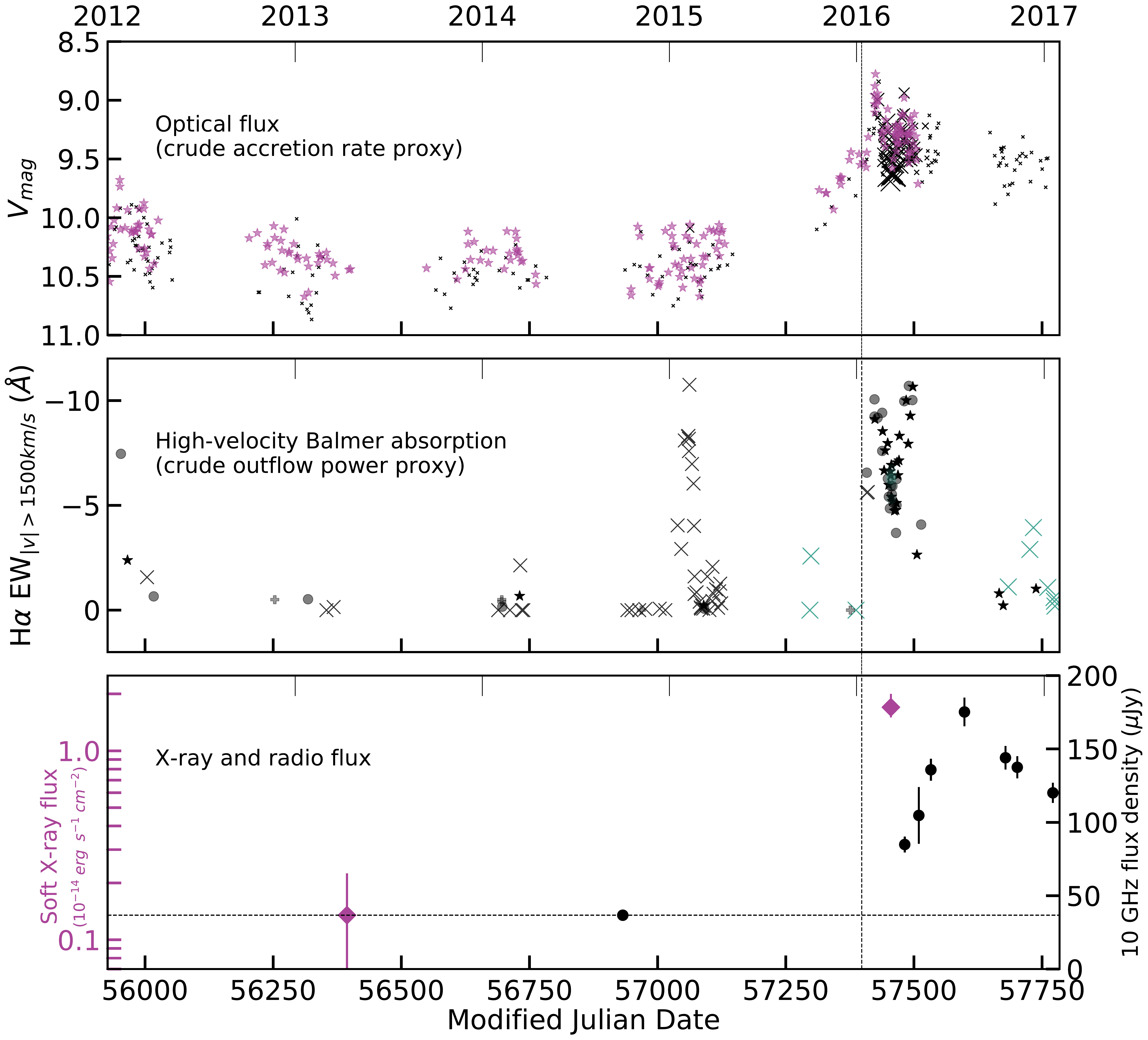}
\caption{Multi-wavelength variability in MWC 560 from 2012 January 1 through 2017 January 31 illustrates the sudden onset of the 2016 outflow fast state following a slow year-long optical rise. The abrupt jump in high-velocity optical absorption strength coincided with a strengthening of soft X-ray flux and the initiation of a linear rise in radio flux density. Also notable is the brief wind burst observed during optical quiescence in 2015 (discussed in \autoref{4.3}), immediately preceding the slow optical flux rise.\protect\\ {\bf Top panel:} V band photometry from the AAVSO \citep[grey X points;][]{Kafka2017} and \citet{Munari2016} (purple star points), a crude proxy for accretion rate through the visual-emitting disc. The AAVSO data are drawn in 1 day bins (taking the median in linear flux units), and the size of the cross is proportional to the number of contributing observations ranging from 1 to hundreds; the \citet{Munari2016} data are not binned. Blank regions are between optical observing seasons.\protect\\
{\bf 2nd panel:} Equivalent width of the H$\alpha$ absorption trough from blue-shifts faster than -1500 km s$^{-1}$, from R=18000--20000 Echelle spectra (black star points), R=1300 spectra taken at Asiago and L'Aquilla (grey circle points) and at Liverpool (blue-green circle points). These data are supplemented in poorly sampled time periods by spectra with a variety of resolutions obtained at the SAO by E. Barsukova and V. Goranskij (2017, private communication; grey plus points), obtained in Tata (blue-green X points), and obtained by volunteers published in the ARAS database (grey X points). Spectra were smoothed to a common resolution before measuring the equivalent width. Blank regions are between optical observing seasons.\protect\\
{\bf 3rd panel, left axis (purple):} Intrinsic soft (plasma temperature constrained to between 0.1 and 1 keV) X-ray outflow-shock component flux (purple diamond points) from Chandra in 2016 and archival XMM data in 2013. \protect\\  
{\bf 3rd panel, right axis (black):} Radio flux density at 9.8 GHz from the VLA (black circle points). \protect\\ The vertical dotted line at 2016 January 11 marks the sudden onset of high-velocity absorption, approximated as the midpoint between the last pre-velocity-jump and first post-velocity-jump spectra.} \label{fig2}
\end{figure*}

\begin{figure}
\centerline{\hspace*{-0.25in}\includegraphics[width=3.0in]{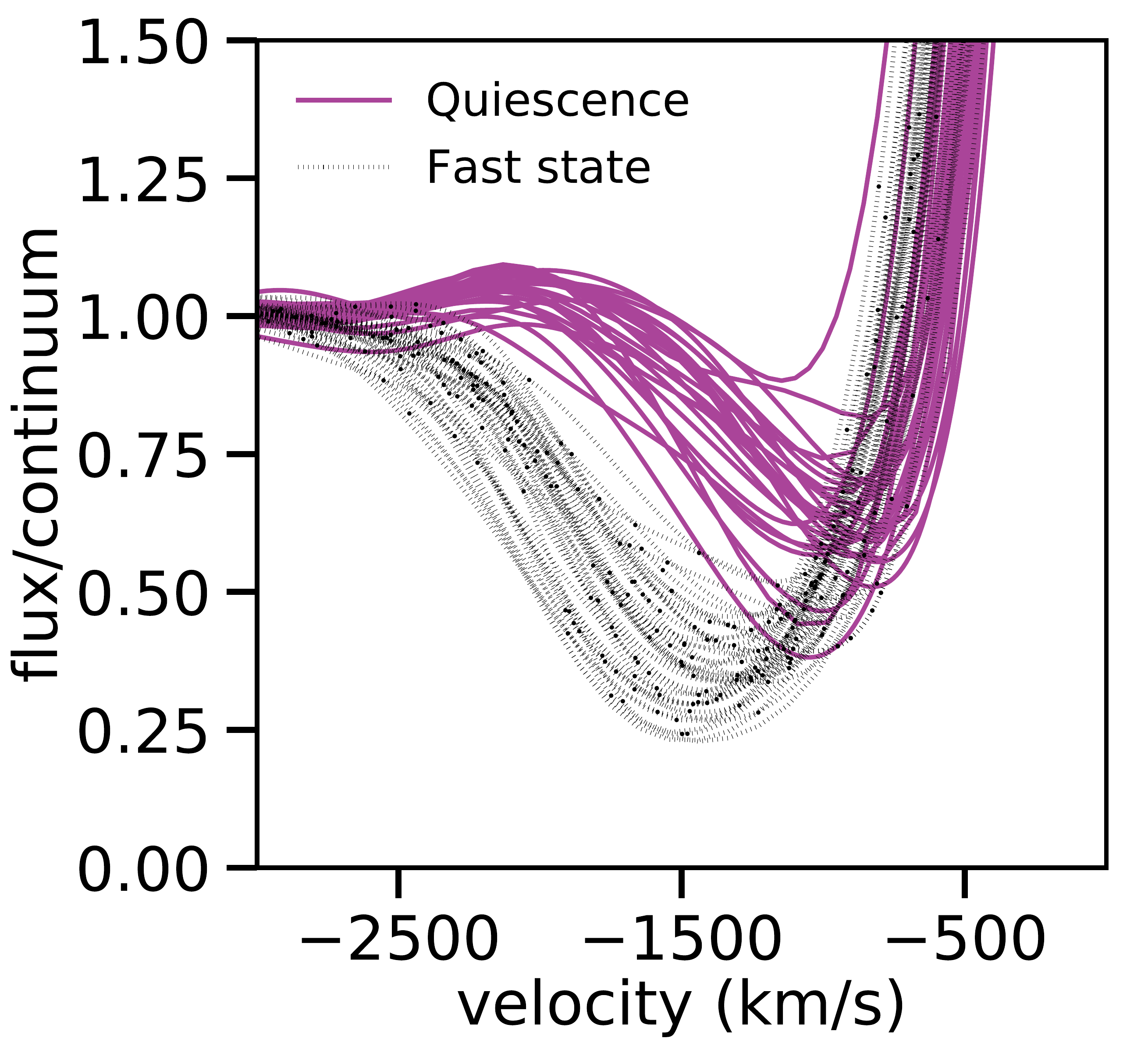}}
\caption{All H$\alpha$ velocity spectra obtained 2016 January through mid-April during the outflow fast state (solid black lines) or obtained during the 2012--2015 quiescence (solid purple lines; excluding only a 1-month wind burst starting in 2015 January, which is discussed in \autoref{4.3}), smoothed to a common resolution of R=450.} \label{optical_beforeafter}
\end{figure}

\section{Data Analysis and Results} 
\label{results}

\autoref{fig2} shows that at the peak of a year-long rise in optical flux tracked by \citet{Munari2016} and the AAVSO \citep{Kafka2017}, the 2016 outflow fast state began abruptly: Balmer absorption-line velocities suddenly doubled in January 2016 (\autoref{opticalabs}), radio emissions began rising by 20$\mu$Jy/month (\autoref{resultsradio}), and soft X-rays strengthened by $\sim10$ times (\autoref{resultsxrays}). The fast state ended by the time that radio flux began gradually descending in July 2016 as Balmer absorption velocities decreased; still, Balmer absorption maximum velocities remained higher than 1500 km s$^{-1}$ even at the end of 2016, and as usual, portions of their profile were consistently saturated up to at least H$\gamma$. The NUV iron curtain absorption remained optically thick throughout our observations (\autoref{uvspectrasec}). Optical/NUV flickering were also persistent throughout (\autoref{flickering}).

\subsection{Optical absorption} \label{opticalabs}

Sometime after the last day of 2015 and before 2016 January 21, the maximum velocity of MWC 560's broad, blue-shifted Balmer absorption lines doubled from below 1500 km s$^{-1}$ into the 2500--3000 km s$^{-1}$ range. They stayed in this fast state until at least mid-April. The timeline is illustrated in \autoref{fig2}. A 1-month velocity burst in 2015 January--February and the longer outflow fast state in 2016 stand out in this figure with large equivalent widths of high-velocity Balmer absorption.

H$\alpha$ velocity profiles in \autoref{optical_beforeafter} show that 2016 January through mid-April saw {\it consistently} higher velocities than the preceding four years of quiescence, except for the 1-month burst of high-velocities in 2015 January--February. In the months following 2016 April, the Balmer velocities gradually decreased, leading to profiles (not plotted) intermediate between typical quiescent and fast state profiles. Large sections of the line profiles remained saturated in high-resolution spectra throughout, for Balmer transitions up to at least H$\gamma$ (and H$\delta$ when included in the wavelength coverage). At no point during 2016 did maximum Balmer velocities drop below 1500 km s$^{-1}$ in high-resolution spectra.

The appearance of high-velocity Balmer absorption in 2016 January was abrupt, occurring entirely during a 3-week gap in observations. The depth of the absorption line relative to the continuum {\it decreased} between 2015 Mar 8 and 2015 Oct 4, and between 2015 Oct 4 and 2015 Dec 31, which we attribute only to a rise in the underlying emission line. The absorption optical depth and velocity did not change in this period, even as late as 2015 December 31.

We supplemented our data in \autoref{fig2} and \autoref{optical_beforeafter} with spectra from other sources, but only in time periods which were poorly sampled by our data (before and after the 2016 outflow fast state). 47 spectra were obtained from the Astronomical Ring for Access to Spectroscopy (ARAS) public database\footnote{\url{http://www.astrosurf.com/aras/Aras_DataBase/Symbiotics/V694Mon.htm}}, comprising all spectra before or after the 2016 outflow fast state and covering H$\alpha$ with a large enough wavelength range for continuum placement. A further 5 spectra were obtained at the Special Astrophysical Observatory (SAO) by E. Barsukova and V. Goranskij (2017, private communication). We focus on H$\alpha$ because it allows the best time domain coverage in available data, by far, and a continuum that is easy to place. Individually examining the higher-order Balmer lines, of which we have poorer temporal coverage but in which the RG contributes less to the continuum, we found the same variability patterns as for H$\alpha$.

To obtain the optical absorption plots and equivalent width calculations discussed in this section, we first smoothed all spectra to a common resolving power R=450, then normalized the flux to the continuum in all spectra as measured by the median of the flux -4000 -- -3000 km s$^{-1}$ blueward of the rest wavelength of H$\alpha$ (with rare exceptions on narrow-band spectra that do not include this wavelength range, in which case a continuum region was manually obtained). Then we measured the equivalent width between -1500 and -3000 km s$^{-1}$, excluding wavelengths with flux higher than the continuum. Lower-velocity absorption below -1500 km s$^{-1}$ was ignored in order to minimize contamination by the variable broad H$\alpha$ emission line. The most variable part of the H$\alpha$ absorption line throughout the 2012--2016 period was on the high-velocity end---and the Balmer absorption lines were always contiguous with their associated emission lines---so this method accurately reflects changes in the whole line. We closely examined the spectra individually, validating that this method did not exaggerate the abruptness of line variability.

In Appendix~\ref{app3}, we demonstrate a correlation between high-velocity Balmer absorption strength and optical/NUV flux on week time-scales throughout the 2016 outflow fast state, as both varied together in a narrow range around their maxima. High-resolution velocity profiles of H$\alpha$, H$\beta$, and \ion{Fe}{ii} in the fast state are also presented in that appendix, along with further details on the spectral smoothing used in \autoref{fig2} and  \autoref{optical_beforeafter}.

\subsection{Radio rise} \label{resultsradio}

\begin{table}
\begin{threeparttable}
\begin{tabular}{clcc}
\hline
Bands & Epoch & $\alpha$ & $\sigma_{\alpha}$\\
\hline
SX (3.1 to 9.8 GHz) & 2016 May 24 & 0.05 & 0.07 \\
 & 2016 Jul 29 & 0.06 & 0.07 \\
 & 2016 Oct 12 & -0.10 & 0.07 \\
 & 2016 Nov 09 & 0.02 & 0.08 \\
 & 2017 Jan 18 & 0.06 & 0.08 \\
 \\
SKa (3.1 to 33 GHz) & 2016 Jul 29 & 0.05 & 0.06 \\
 & 2016 Oct 12,17\tnote{a} & -0.03 & 0.06 \\
 & 2016 Nov 09 & 0.08 & 0.06 \\
 & 2017 Jan 18 & 0.05 & 0.06 \\
 \\
XKa (9.8 to 33 GHz) & 2016 Jul 29 & 0.04 & 0.12 \\
 & 2016 Oct 12,17\tnote{a} & 0.05 & 0.11 \\
 & 2016 Nov 09 & 0.14 & 0.11 \\
 & 2017 Jan 18 & 0.05 & 0.12  \\
\hline
\end{tabular}
\caption{Radio spectral indices. Spectral index is defined such that $F\propto\nu^{\alpha}$ between the two given bands; approximate central frequencies of the bands are listed for convenience. $\sigma_{\alpha}$ is the uncertainty in the spectral index $\alpha$ propagated from the total uncertainty $\sigma_{tot}$ in flux from \autoref{tableradio}. \label{alpha}}
\begin{tablenotes}
\item[a] \footnotesize Epoch 2016 Oct 12,17 denotes $\alpha$ for SKa and XKa wherein the Ka band observation preceded the S and X band observations by 5 days.
\end{tablenotes}
\end{threeparttable}
\end{table}

The onset of the 2016 outflow fast state coincided with a rapid, roughly 20$\mu$Jy/month rise in flat-spectrum radio emissions up to a maximum of 175 $\pm$ 10 $\mu$Jy at 9.8 GHz on 2016 July 29, about 5 times brighter than the flux density observed in the only prior radio detection of this system on 2014 October 2\footnote{A non-detection was incorrectly reported for the 2014 October 2 observation in \citet{Weston2016} and \citet{Lucy2016a}, and corrected by the erratum \citet{Lucy2017}.}. We plot the 9.8 GHz flux density measurements as a function of time in \autoref{fig2}. The flux densities observed at 3.1, 9.8, and 33.1 GHz are tabulated in \autoref{tableradio}.

The radio spectrum was always flat between all observed bands, with spectral index $\alpha$ ($F\propto\nu^{\alpha}$, F=flux and $\nu$=frequency) between -0.1 and 0.14 (modulo uncertainties $\sim0.1$); \autoref{alpha} lists these measurements. In all cases except the SKa and XKa indices for October, the observations used to calculate $\alpha$ were obtained nearly simultaneously, within 3 hours of each other. In all cases, the in-band spectral indices are also consistent with being flat within large uncertainties.

No intra-observation variability was apparent from the visibilities; in particular, we checked for X-band variability within our brightest observation (2016 July 29) by modelling all other sources in the field, subtracting the model using {\it uvsub} in CASA, and examining the residuals in the visibilities. We also imaged that observation in four equal time sections. The flux of MWC 560 in this observation was constant within uncertainties (which were 3--4 times larger in the shorter images than for the whole observation). 

We checked for polarized emission for the 2016 July 29 and October 17 observations using S and X band data, and found none. MWC 560 was not detected in Stokes Q and U, in which the noise was only slightly higher than in Stokes I.

\begin{figure}
\centerline{\includegraphics[width=3.5in]{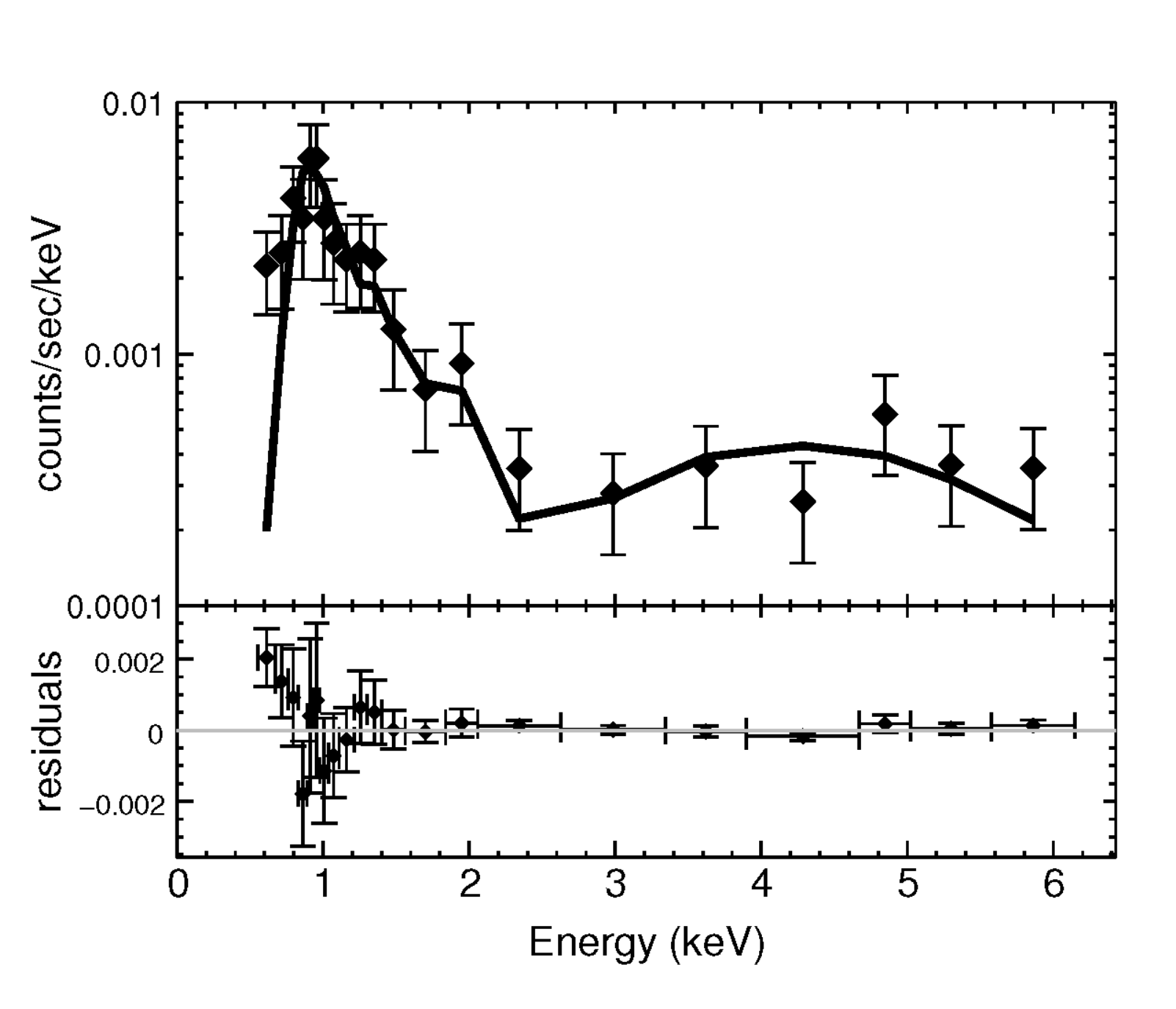}}
\caption{The background-subtracted X-ray spectrum obtained with Chandra on 2016 March 8--9, displayed with grouping to bins of at least 10 counts and fit with the 2-component soft$+$hard fit described in the text and \autoref{tablexray}. The fit itself was performed on the spectrum grouped to bins of at least 20 counts. The vertical axis is logarithmic for the spectrum and linear for the residuals.} \label{xrayspectra}
\end{figure}

\begin{table}
\begin{threeparttable}
\resizebox{0.48 \textwidth}{!}{
\begin{tabular}{lllll}
\hline
Component & Parameter & Best fit & -2$\sigma$ & +2$\sigma$\\
\hline
soft & kT (keV) & 0.77 & -0.66 & +0.22\\
 & norm & 2.0E-05 & -1.2E-05 & +0.036\\
 & nH (10$^{22}$ cm$^{-2}$) & 0.46 & -0.33 & +0.64\\
 \\
hard & kT (keV) & 11.26 & frozen & \\
 & norm & 8.9E-05 & -6.7E-05 & +1.39E-4\\
 & nH (10$^{22}$ cm$^{-2}$) & 13.48 & -12.74 & +19.39\\
 \hline
\end{tabular}
}
\caption{X-ray model for Chandra 2016 March 8-9 epoch. Both components with temperature kT and normalization norm were modelled using the CIAO {\it Sherpa} clones of the XSPEC {\it apec} diffuse collisionally-ionized plasma model with solar abundance ({\it angr}) and redshift 0, each absorbed by their own {\it wabs} model with hydrogen column nH. \label{tablexray}} 
\end{threeparttable}
\end{table}

\subsection{Strengthened soft X-ray flux} \label{resultsxrays}

Our Chandra spectrum on 2016 March 8--9 showed the emergence of a soft X-ray component about $10$ times stronger than during the 2007 and 2013 X-ray epochs. The hard X-ray component remained as weak as it was in 2013. The source aperture contained $254$ total source counts summed over the pair of exposures, amounting to $5.15 \pm 0.3 (1\sigma) \times 10^{-3}$ source counts s$^{-1}$. Our data and best fit model are plotted in \autoref{xrayspectra}, and the best fit parameters are listed in \autoref{tablexray}.

\begin{figure}
\centerline{\includegraphics[width=4.0in]{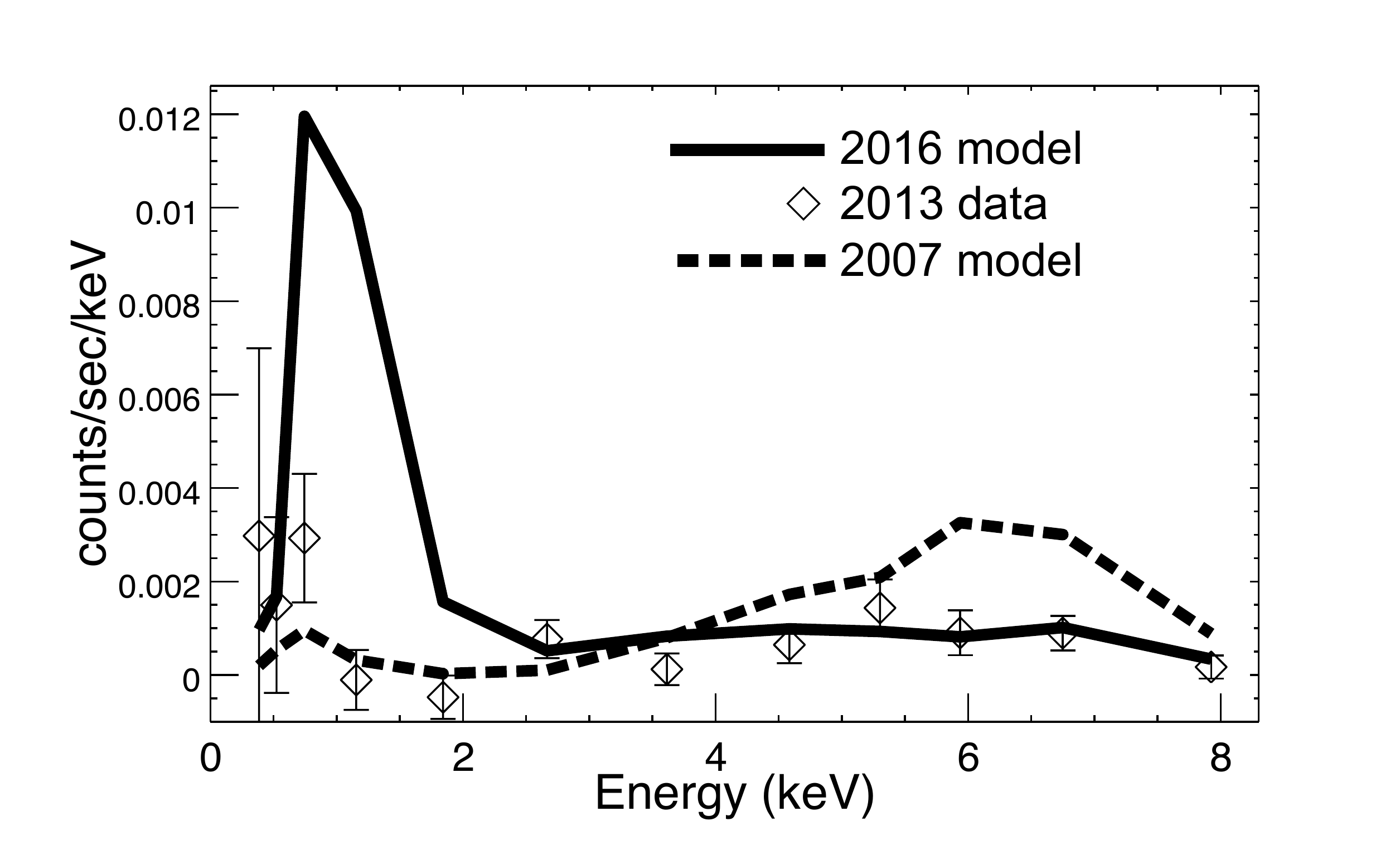}}
\caption{X-rays in 2007 (dotted line), 2013 (data points), and 2016 (solid line). Only the soft component was bright in 2016, only the hard component was bright in 2007, and neither component was bright in 2013. We demonstrate this point using the \citet{Stute2009} model obtained for their 2007 September 27 XMM spectra, the 2013 April 12 XMM PN X-ray spectrum with background subtracted, and the 2016 March 8-9 model shown in \autoref{tablexray} and \autoref{xrayspectra}.} \label{vs2013}
\end{figure}

The soft component was weak in 2007 and 2013 and strong in 2016, while the hard component was strong in 2007 and weak in 2013 and 2016. This result is obvious from \autoref{vs2013}, which plots the 2007 September 27 best fit model reported by \citet{Stute2009}, our background-subtracted reduction of the 2013 April 12 XMM PN data, and our 2016 best fit model.

Following the MWC 560 models used by \citet{Stute2009} and the general symbiotic star models used by \citet{Luna2013}, we obtained our 2016 model by fitting both the hard and soft components with separately-absorbed (photoelectric absorption with Wisconsin cross-sections), collisionally-ionized, optically-thin diffuse plasma emission spectra. These models were inspired by \citet{Patterson1985b} for emission of a hard component by the extended halo  of a WD accretion disc's boundary layer (BL), and \citet{Muerset1997} for emission of a soft ($\beta$-type) component by colliding winds; the soft component is almost certainly not supersoft. In CIAO's {\it Sherpa}, we grouped counts from our co-added Chandra spectra into bins of 20 source-region counts, and fit the spectrum over the range for which the Chandra energy scale is calibrated (0.277--9.886 keV) with a \textsc{(wabs\textsubscript{soft}$\times$apec\textsubscript{soft})$+$(wabs\textsubscript{hard}$\times$apec\textsubscript{hard})} model. The small flux of the hard component in our spectrum required us to fix its temperature to the 11.26 keV best fit obtained during its brightest state, the 2007 epoch \citep{Stute2009}.

The soft component, which has a temperature somewhere between 0.1 and 1 keV and remained roughly constant between 2007 and 2013, had an observed, absorbed flux at least 7 times brighter in 2016 (best fit 13 times) than in the prior epochs. This strengthening of the soft X-ray flux is plotted in \autoref{fig2}. The 2016 observed flux from the soft component model was $1.7_{-0.2}^{+0.3}\times10^{-14}$ erg s$^{-1}$ cm$^{-2}$, which also acts as a lower limit on the intrinsic, unabsorbed flux. The upper limit on the unabsorbed flux is poorly constrained\footnote{This is because the normalization of the soft component flux is degenerate with the absorbing hydrogen column. Absorption mainly affects the peak and low-energy flank of the observed soft component flux, so at high temperatures in the allowed parameter space, the high-energy flank places an upper limit to the normalization. At lower temperatures, however, where the model's high-energy flank is weak relative to the rest of the profile, flux normalization and absorbing column can be freely increased together without much affecting the modelled spectrum.} at $\lesssim3\times10^{-11}$ erg s$^{-1}$ cm$^{-2}$. These correspond to luminosities of $1.3^{+0.2}_{-0.2}\times10^{31}$ erg s$^{-1}$ (d/2.5kpc$^{2})$ and 
$\lesssim2\times10^{34}$ erg s$^{-1}$ (d/2.5kpc$^{2})$, respectively.

The hard component, which by 2013 had dimmed to $0.3_{-0.3}^{+0.2}$ times its 2007 observed flux (i.e., dimmed to $\leq$1$\times10^{-13}$ erg s$^{-1}$ cm$^{-2}$), remained dim in 2016. No statistically significant change occurred in this component between 2013 and 2016.

In Appendix~\ref{app4}, we demonstrate that there is no evidence for X-ray variability {\it during} the 2016 outflow fast state; however, the source was not bright enough for strong constraints. In Appendix~\ref{app5}, we report the details of our fitting routines and the calculation of inter-epoch variability constraints. In Appendix~\ref{app6}, we demonstrate that optical loading definitely does not affect our Chandra observations or our main conclusions, and probably does not affect our Swift XRT observations.

\begin{figure}
\centerline{\includegraphics[width=3.5in]{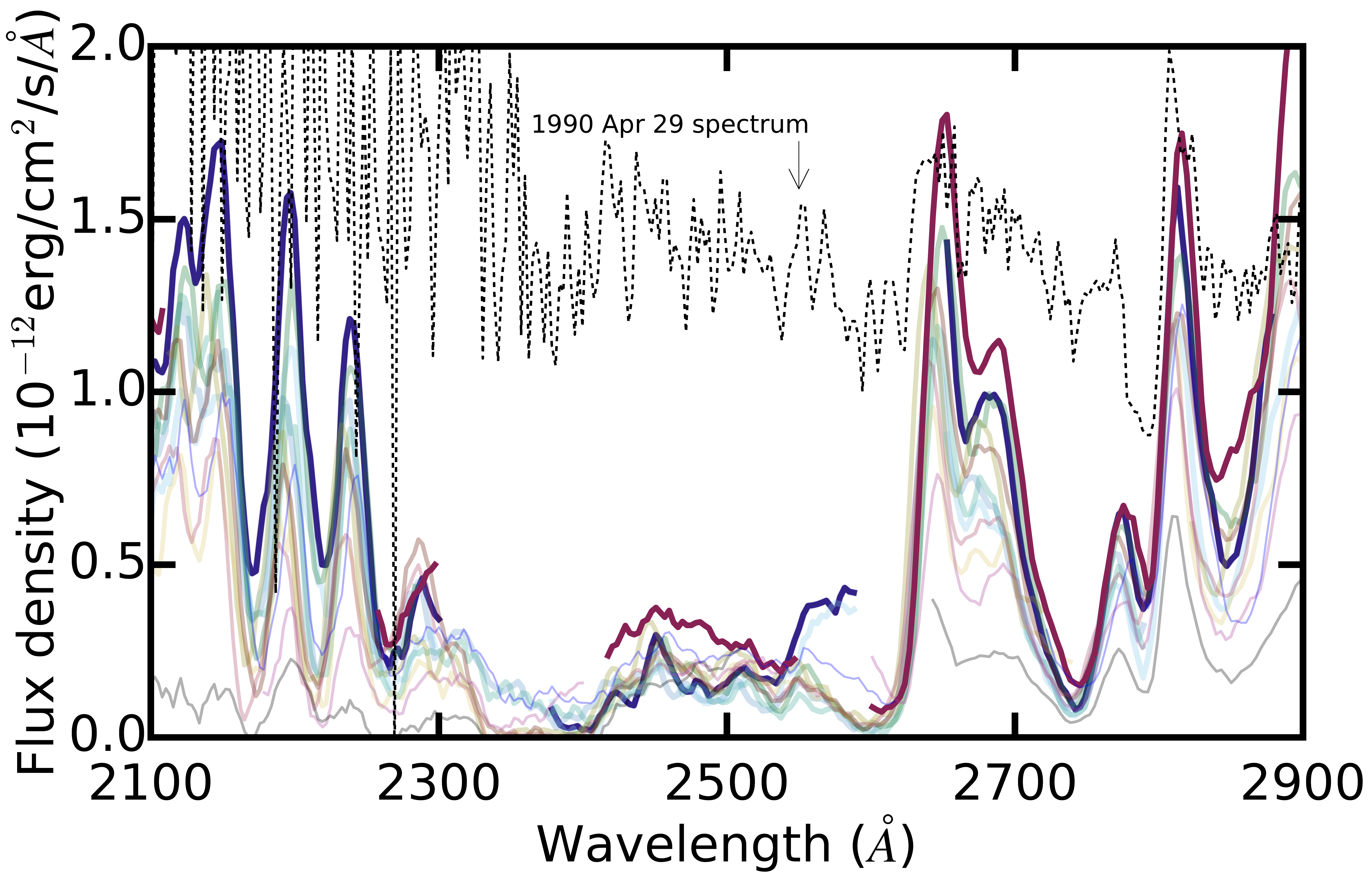}}
\caption{The Swift UV spectra (obtained between 2016 March and June; all solid lines) demonstrate substantial variability in normalization but not in shape, and signify a persistent iron curtain absorbing on a varying continuum throughout the 2016 observations. In particular, the varying flux near 2650\AA\ and at the \ion{Mg}{ii}$\lambda$2800\AA\ emission line suggest that the varying normalization cannot be attributed to varying opacity. The 1990 April 29 spectrum (dashed black line) is included for comparison, and approximates the underlying continuum at the NUV flux maxima of 2016 March 2 (dark blue line) and 2016 April 4 (dark red line). Omitted from this plot are all regions in individual spectra where image inspection suggested zeroth-order contamination, as well as the three spectra noted in \autoref{tableuv} as having nearby bright first order flux or suffering from underestimated flux due to loss of lock. All spectra were de-reddened by E(B-V)=0.15 (\citealt{Schmid2001}, and our Appendix~\ref{app1}).} \label{uvspectra}
\end{figure}

\begin{figure}
\centerline{\hspace*{-0.21in}\includegraphics[width=3.5in]{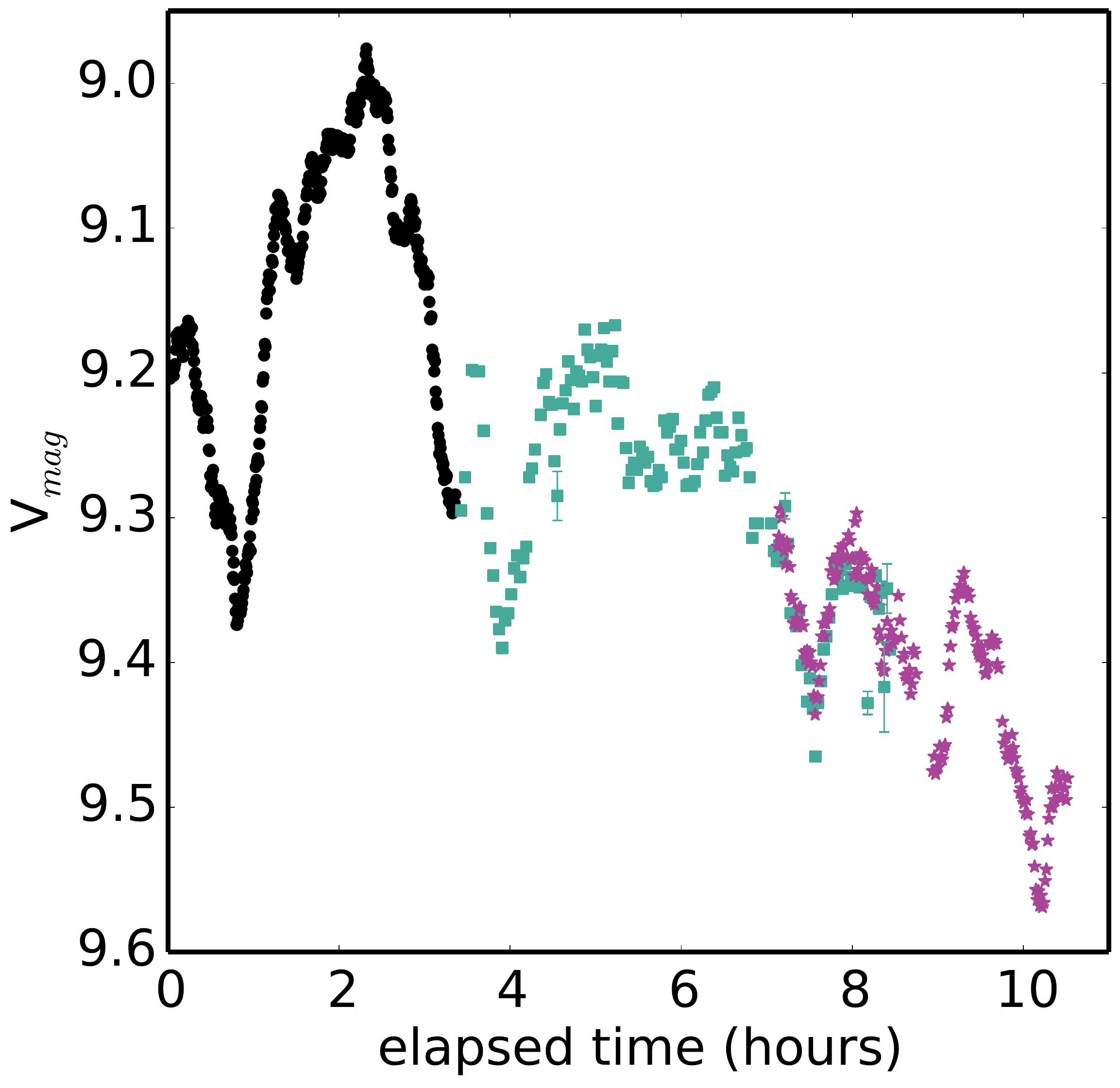}}
\caption{Johnson V band photometry starting at 2016 March 1 at 19:55 UT (MJD 57448.83) replicating data from the AAVSO \citep{Kafka2017} and showing a typical example of the optical flickering that remained persistent throughout the 2016 outflow fast state. Data are colour-coded by observer:
Teofilo Arranz (black circles), Gary Walker (blue-green squares), and Geoffrey Stone (purple stars). Statistical uncertainties submitted by the observer are on the scale of the data points except where error margins are plotted.} \label{vflick}
\end{figure}

\subsection{Persistent ultraviolet absorption} \label{uvspectrasec}

Our NUV spectra, which span 2016 March 29 to 2016 June 1, show that the ``iron curtain'' of overlapping absorption troughs from \ion{Fe}{ii} and similar ions \citep{Michalitsianos1991,Lucy2018} remained optically thick throughout the 2016 outflow fast state. These overlapping broad lines absorb on the continuum and leave behind a pseudo-continuum that looks like, but is not, an emission spectrum. 

\autoref{uvspectra} shows that our 14 usable Swift UV grism spectra keep a roughly constant shape, with the differences between them best explained as a simple scaling factor applied uniformly to the whole NUV spectrum; the variability is in the underlying continuum, not in the absorption. Comparing to MWC 560's spectral morphologies from prior epochs reported in \citet{Bond1984} and \citet{Michalitsianos1991} and also categorized in \citet{Lucy2018}, the 2016 spectra are a reasonable match to the 1990 March 14 spectral morphology, and an even better match to the 1991 September 29 spectral morphology due to additional absorption between 2400-2550\AA, which we ascribe to Fe$^{+}$ lines with an upper level of 2.7 eV.

We further infer that the underlying continuum being absorbed upon at the NUV flux maxima in 2016, March 2 and April 4, is very similar to the spectrum observed by \citet{Michalitsianos1991} on 1990 April 29, which is largely unabsorbed (excepting \ion{Mg}{ii}). We retrieved this spectrum ({\it International Ultraviolet Observer} LWP17832, R$\approx$200-350 comparable to Swift) from the Mikulski Archive for Space Telescopes (MAST), and include it for reference in \autoref{uvspectra}.

\begin{figure}
\centerline{\includegraphics[width=3.5in]{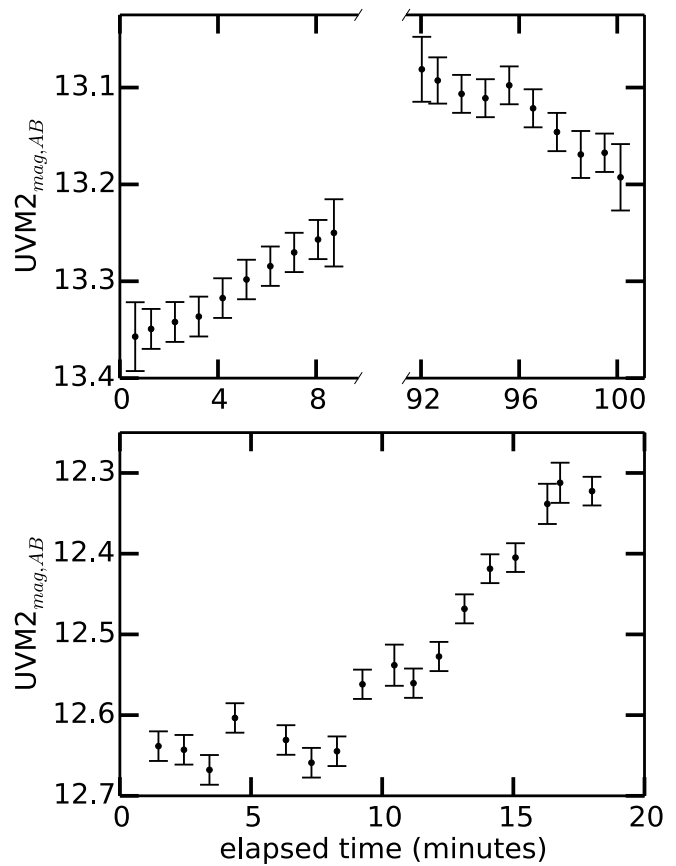}}
\caption{Example light curves from our Swift UVOT observations. These changes in NUV flux, up to 40\% on a time-scale of minutes, are consistent with flickering from the accretion disc. Swift UVM2 AB magnitude is plotted against minutes since 2016 June 1 at about 18:23 UTC (MJD$=57540.7661081$, top panel, with a break in the time axis during which observations were interrupted) and 2016 March 5 at about 5:39 UTC (MJD$=57452.235247$, bottom panel).} \label{UVflick}
\end{figure}

\subsection{Persistent optical+NUV flickering}
\label{flickering}

The rapid flickering in the optical and NUV that almost always characterizes the MWC 560 system \citep[e.g.,][]{Bond1984,Tomov1996,Zamanov2011a,Zamanov2011b} persisted throughout the 2016 outflow fast state. Densely-sampled AAVSO V-band observations covered February through April, our Swift UVM2 observations covered late-February into June, and strong variability on short time-scales was reliably observed throughout.

V-band magnitudes typically varied by at least 0.1 mag per 15--30 minutes, and the light curves strongly resembled past periods of strong flickering \citep[comparable, e.g., to the bottom panels of Figure 1 in][]{Zamanov2011a}. For example, our \autoref{vflick} reproduces a characteristic light curve from the AAVSO \citep{Kafka2017} on 2016 March 1--2, continuous over 11 hours thanks to volunteers observing at widely separated geographic locations. To verify this result, we confirmed that there was no systematic relationship between flux and airmass in these data. 

The expected NUV counterpart to MWC 560's optical flickering is observationally confirmed here for the first time; the most robust examples are shown in \autoref{UVflick}. Swift UVM2 ($\lambda_{cen}$=2246\AA, FWHM=513\AA) magnitude varied by up to 0.4 magnitudes peak-to-peak over time-scales as short as 10 minutes, consistent with flickering. Longer time-scale NUV flux variability is discussed and plotted in Appendix~\ref{app3}.

It is virtually certain that the rapid UV variability in \autoref{UVflick}, and most of the remaining UV variability included in Appendix~\ref{app3}, is real and not a systematic error. The reference star, TYC 5396-570-1 (AB UVM2 $\approx$15.8 magnitudes), was much less variable than MWC 560 on all time-scales; less than 2\% of its light curve deviated by more than 3$\sigma$ (0.15 magnitudes) from the reference star's median magnitude. In particular, the MWC 560 light curves in \autoref{UVflick} were accompanied by stability in the reference star, with less than 20\% of the data deviating from the local median by more than 1$\sigma$, and no noticeable evidence of trends in the direction of the MWC 560 variability. Furthermore, the UVOT instrument is known to be very stable, and has been shown by the Swift team to vary by less than 0.1 magnitudes in the brightness measured for a WD calibration source placed at one location on the detector over the course of weeks.

\section{Discussion} \label{discussion}

At the peak of a year-long rise in accretion rate through the visual-emitting disc, the dense accretion disc outflow abruptly jumped in power in January 2016. The dense outflow remained fast and stable for several months (\autoref{4.1.1}), steadily feeding a lower-density region of radio-emitting gas (\autoref{4.1.2}), and shocks at the collision of high-velocity and low-velocity components of the dense outflow began radiating a large soft X-ray flux (\autoref{4.1.3}). The inner accretion disc remained intact throughout (\autoref{4.2}). Properties of the system during this outflow fast state, calculated throughout these sections, are listed in \autoref{tablevalues}. The stability of the outflow and the disc in 2016 stands in marked contrast to the instability and disruption of both outflow and disc in 1990, even though the 2016 and 1990 brightening events reached a similar peak accretion rate. We ascribe 2016's stability to the regulatory and stabilizing effect of the outflow on the accretion disc during both a short high-velocity burst in early-2015 and during the 2016 outflow fast state (\autoref{4.3}).

\begin{table*}
\begin{threeparttable}
\begin{tabular}{ll}
\hline
Density of absorption-line region: &  $\gtrsim$10$^{6.5}$ cm$^{-3}$\\
Column density of outflow: & $\gtrsim$10$^{23}$ cm$^{-2}$\\
Maximum outflow velocity: & 2500--3000 km s$^{-1}$\\
Strong-shock velocity for soft X-rays: & 300--900 km s$^{-1}$\\
Radio emission mechanism: & optically-thin thermal\\
Density of radio-emitting region: & $\lesssim$10$^{5.5}$ (2.5 kpc/d)$^{1/2}$ cm$^{-3}$\\
Outflow radius for radio emissions: & $\gtrsim$60 (d/2.5 kpc) au\\
Mass-outflow rate for radio emissions:\tnote{*} & $\gtrsim$10$^{-6}$ (d/2.5 kpc)$^{5/2}$~M$_{\odot}$~yr$^{-1}$\\
Inner accretion disc: & intact with persistent optical/UV flickering\\
Accretion-disc bolometric luminosity peak: & 1800 (d/2.5 kpc)$^{2}$ L$_{\odot}$\\
Accretion rate peak:\tnote{$\dagger$} & 6$\times$10$^{-7}$ (d/2.5 kpc)$^{2}$ (R / 0.01 R$_{\odot}$) (0.9 M$_{\odot}$/ M)~M$_{\odot}$ yr$^{-1}$\\
\hline
\end{tabular}
\begin{tablenotes}
\caption{Physical properties of the outflow fast state (Jan-Jul 2016).} \label{tablevalues}
\item[*] \footnotesize{For simplified uniform-density model, assuming no clumping: consistency check only.}
\item[$\dagger$] \footnotesize{If half of the accretion luminosity is emitted by the boundary layer in the extreme UV and not re-emitted, as assumed for MWC~560 by Schmid et al. (2001), we would obtain twice the accretion rate, about $1\times10^{-6}$ M$_{\odot}$~yr$^{-1}$.}

\end{tablenotes}
\end{threeparttable}
\end{table*}

\subsection{The abrupt onset of a stable fast state}\label{4.1}

\subsubsection{The dense, stable, fast absorption-line outflow}\label{4.1.1}

The best explanation for the jump in Balmer absorption maximum velocities between 2015 December 31 and 2016 January 21 (\autoref{opticalabs}) is the sudden appearance of high-velocity material in MWC 560's dense outflow. The appearance of high-velocity Balmer absorption occurred abruptly only towards the peak of a year-long, smooth rise in optical flux, without any evidence of a commensurately abrupt change in photoionizing flux, so it is unlikely to have been due to a photoionization effect. The new fast flow remained until at least mid-April, sometime after which it began very gradually slowing down throughout the year.

{\it After} the abrupt velocity jump, the outflow was remarkably stable and predictable throughout its 2016 fast state. High-velocity Balmer absorption, which was always present during the fast state, varied slightly on week time-scales in time with the varying optical and NUV flux; this correlation between optical depth and broadband flux was likely attributable to a photoionization effect (Appendix~\ref{app3}).

As usual in MWC 560, the absorption-line outflow was dense. The saturation to black of large portions of the H$\gamma$ absorption trough (\autoref{opticalabs}; $\tau\gtrapprox1.6$) required at minimum a density n$_{H}\gtrsim10^{6.5}$ cm$^{-3}$ and a column density N$_{H}\gtrsim10^{23}$ cm$^{-2}$, for any ionization parameter and assuming turbulent velocities $\gtrsim100$~km~s$^{-1}$ \citep{Hamann2019,Williams2017}. The persistence of the NUV iron curtain (\autoref{uvspectrasec}) is also consistent with continually high densities, although it is not as restrictive: n$_{H}\gtrsim10^{5}$ cm$^{-3}$ for excited states up to 1.1 eV above the ground state \citep{Lucy2014}, which we certainly observe in MWC 560. We also tentatively identified \ion{Fe}{ii} transitions from 2.7 eV above the ground state; by comparison, only some quasars with broad Balmer absorption also feature these extremely high-excitation lines, so they may impose stricter lower limits on the density in photoionization modelling \citep{Aoki2010}.

\afterpage{
\begin{figure*}
\centering{\includegraphics[width=7in]{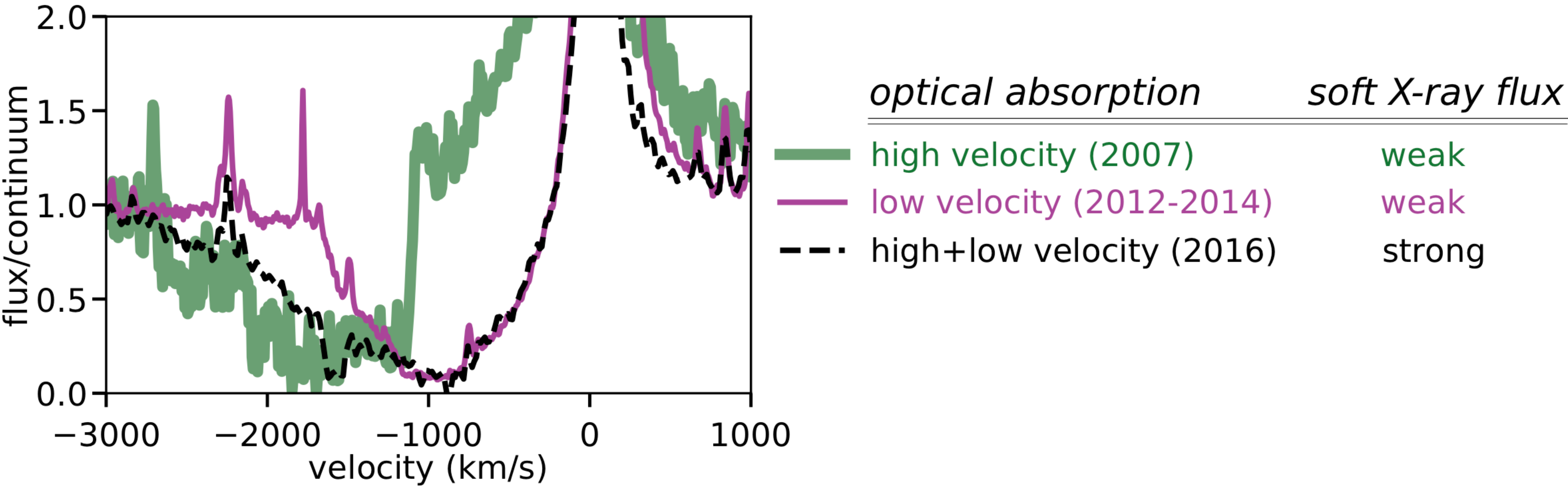}}
\caption{High-resolution H$\beta$ velocity spectra characterize the state of the optical absorption during the three X-ray epochs. A strong soft X-ray flux is observed only when both the low-velocity and high-velocity outflow components are simultaneously present, suggesting that the soft X-ray $\beta$ emission may be emitted by a shock at the collision between fast and slow absorbers. The optical spectra were obtained on 2007 October 30 (thick green line; a month after the 2007 Sep 27 X-ray epoch), 2012 Feb 8 (thin purple line; the closest high-resolution match, when smoothed, to a low-resolution 2013 Jan 26 spectrum obtained 3 months before the 2013 Apr 12 X-ray epoch), and 2016 Mar 8 (dashed black line; the first day of our 2016 X-ray epoch).} \label{fignew}
\end{figure*}
}

\subsubsection{Steadily feeding a lower-density radio emission region}\label{4.1.2}

The radio data provide further evidence that January 2016 marked a jump in outflow power and the onset of a stable fast state. In this section, we argue that the radio emissions originated as optically-thin thermal emission in a region of low-density ($\lesssim10^{5.5}$\,cm$^{-3}$) gas being steadily fed by the fast, denser absorption-line outflow. 

The flat-spectrum radio flux density doubled in the 4 months between 2016 April 4 and July 29, growing even as the optical/NUV flux varied up and down. The slope of this radio rise is consistent with linear growth starting at its quiescent-state value in January 2016 (inferred to be 37 $\mu$Jy per the 2014 observation), roughly corresponding to the onset of fast optical absorption. Further supporting a fast-outflow origin for the radio emissions, the H$\beta$ absorption profile for 2014 October 30, obtained less than a month after the quiescent-state radio flux was measured to be weak, lacks the high-velocity absorption component that appeared in January 2016.

We favour thermal emission as the radio source, in a growing region steadily fed by the stable absorption-line outflow. The radio spectral index $\alpha$ is in the range -0.17 to 0.25 (which includes measurement uncertainties at the edges) and categorically rules out the $\alpha\geqslant0.6$ often observed from many symbiotic star nebulae due to optically-thick thermal bremsstrahlung emission with a radially-dependent photosphere from an asymmetrically-ionized wind \citep{Seaquist1993,Seaquist1990,Seaquist1984,Reynolds1986,Wright1975,Panagia1975}, as well as the $\alpha\approx-0.7$ of optically thin synchrotron emission. Indeed, while symbiotics---especially S-type symbiotics like MWC 560---seem to generally have $\alpha$\textgreater0.6 in quiescence (\citealt{Seaquist1993,Seaquist1990}), flatter indices tend to emerge from jets \citep[e.g.,][]{Weston2016}. An outflow origin is therefore more plausible than any alternative source for the radio emission.

Furthermore, an optically thick / self-absorbed synchrotron emission mechanism, such as is favoured by \citet{Coppejans2016}, \citet{Russell2016}, and \citet{Koerding2008} for their dwarf nova radio outbursts, is not consistent with the brightness temperatures for our data unless the emitting source is extremely small. In particular, the brightness temperature at 10 GHz (where there is no doubt the spectrum is flat) ranges from 0.6--1.3$\times10^{7}$ (1 au/s)$^{2}$ (d/2.5kpc)$^{2}$ K where s is the average of the outflow's major and minor axes, and d is the distance to the system. The kinetic temperature of synchrotron-emitting electrons at this frequency is 1.0$\times10^{13}$ (B / 10$^{-4}$ Gauss)$^{-1/2}$ K where B is the magnetic field strength \citep[e.g.,][]{Condon2016}. The brightness temperature must be on the order of the kinetic temperature of the emitting electrons for synchrotron self-absorption to be significant and the spectrum to flatten \citep[e.g.,][]{Williams1963}. Given the speed of the outflow (travelling roughly 1 au away from the disc per day), it would be difficult to contrive a situation where the emitting region is compact or collimated sufficiently to fulfil this condition. Fulfilling the condition becomes even more difficult (by a factor of about 20) if the spectrum really is flat out to 33 GHz, as it appears to be.

Given our conclusion that the emission is thermal, the flat spectrum suggests an almost fully optically thin emitting region, which places an upper limit on the density and a very rough lower limit on the mass outflow rate. For such a gas, the Rayleigh-Jeans Planck approximation yields

\begin{equation}
F_{\nu} \approx (T_{e}/1200K)( \nu /GHz)^2 ( \theta_{s}/arcsec)^2 \tau
\end{equation}

where $F_{\nu}$ is flux density in mJy,  $\theta_{s}$ is the angular size of the emitting gas in the plane of the sky, the electron temperature $T_{e}$ is set to the gas temperature, and the optical depth is

\begin{equation}
\tau \approx 3.28\times10^{-7}(T_{e}/10^{4}K)^{-1.35}(\nu/GHz)^{-2.1}(n_{e}^{2}s/cm^{-6}pc)
\end{equation}

where s is the size of the emitting gas along the line of sight, for the simple case of uniform gas density equal to the electron density $n_{e}$. For $\tau\lesssim0.1$, $T_{e}\lesssim10^{4}$~K, $F_{\nu}=175~\mu$Jy, and $\nu=$~10~GHz, we obtain an emitting region size scale in the plane of the sky of $\gtrsim6\times10^{-4}$~(d/2.5 kpc)~pc (i.e., $\gtrsim120$ au; as a consistency check, this scale could be reached by a bipolar outflow in $\gtrsim50$~days at 2000~km~s$^{-1}$). Assuming that the size scale in the plane of the sky is comparable to the size scale along the line of sight, we obtain a density $\lesssim10^{5.5}$~(2.5~kpc/d)$^{1/2}$~cm$^{-3}$. As a check on plausible mass outflow rates, we can divide the radio emitting region's mass by the time-scale over which it was formed. If it took about 200~days for the radio emission to reach 175$\mu$Jy (about the time between the emergence of high-velocity optical absorption and the 2016 July 29 radio observation), then a region with this volume and density could be filled by a mass outflow rate $\gtrsim10^{-6}$~(d/2.5 kpc)$^{5/2}$~M$_{\odot}$~yr$^{-1}$, modulo the simplifying assumptions made for this toy model. However, if there is clumping in the outflow, the mass outflow rate could be lower.

\subsubsection{Soft X-rays: shocks at the collision\\of fast and slow absorbers}\label{4.1.3}

\autoref{fignew} suggests that strengthened soft X-rays during the 2016 outflow fast state may have originated in a shock at the collision of the new high-velocity absorbers and the pre-existing low-velocity absorbers. The optical spectrum most coeval to the 2007 X-ray epoch exhibited only high-velocity Balmer absorption, lacking absorption below $\sim$1000~km~s$^{-1}$. Spectra obtained during the 2012-2014 quiescent-outflow periods, including one low-resolution spectrum obtained within 2.5 months of the 2013 X-ray epoch (an excellent match to the echelle spectrum used in \autoref{fignew}, when the latter is smoothed to lower resolution), exhibited only lower-velocity absorption. Soft X-rays were weak in both 2007 and 2013. It was only in the 2016 outflow fast state, when high-velocity and low-velocity absorbers coexisted, that the soft X-ray component brightened by about an order of magnitude or more. 

Our fit to the soft X-ray component observed on 2016 March 8-9 constrained the diffuse, collisionally-ionized gas component temperature to between 0.1 and 1.0 keV. This corresponds to a strong-shock velocity of 300 to 900 km s$^{-1}$, consistent with the differential velocities between the high-velocity and low-velocity absorbers. The soft X-rays emission measure, n$^{2}$V$\sim10^{54}$ (d/2.5 kpc)$^{2}$ cm$^{-3}$, is easy to achieve; assuming a density $10^{7}$ cm$^{-3}$ and a spherical toy model, this corresponds to an emission-region radius $\sim10^{-6}$ pc (i.e., $\sim0.1$ au), which can be traversed by a 2000 km s$^{-1}$ outflow in mere hours. In contrast, the soft X-ray component is likely too hard to be supersoft emission from nuclear burning, and too soft to be boundary-layer emission. We can certainly rule out softening of a previously observed hard X-ray component; in the 2013 epoch there was no comparably bright hard component, and even at the hard component's maximum in 2007 it was dimmer than the 2016 soft component.

\subsection{Intact inner disc}
\label{4.2}

The inner accretion disc of MWC 560 remained intact throughout 2016, and the increase in luminosity was powered by accretion alone without nuclear burning. Variability in optical/NUV photometry corresponds to variability in the accretion rate through the optical/NUV-emitting parts of the disc.

Persistent optical flickering was observed throughout the 2016 outflow fast state (\autoref{flickering}) and in every preceding monitoring project except in 1990 (e.g., \citealt{Bond1984,Tomov1996,Zamanov2011b,Zamanov2011,Zamanov2011a}). Rapid flickering most likely comes from instabilities in an inner accretion disc near the boundary layer, where the instability time-scales are short. While disc-less models are presumably conceivable, to our knowledge there is no evidence for disc-less flickering from accreting WDs \citep[see review in][]{Sokoloski2001}. Hot spots can produce some flickering where the accretion stream hits a disc, but high frequency and large amplitude flickering from accreting WDs is generally caused by the inner disc or BL \citep[see the introduction to][for a review]{Bruch2015}. We also confirmed the expected NUV flickering in MWC 560 for the first time; the NUV spectral morphology remained constant, so the variability we observed must have originated in a variable NUV continuum. \citet{Luna2013} have shown that flickering is suppressed by a flood of invariant light if there is nuclear burning on the WD surface, so MWC 560 must be powered by accretion alone. Indeed, the short-term and long-term stability of the absorption NUV and optical absorption spectra (\autoref{opticalabs},\autoref{uvspectrasec},\autoref{4.1.1}) further disfavors a thermonuclear nova interpretation.

We estimate a 2016 peak (V$\approx$8.8) accretion disc bolometric luminosity of about 1800 (d/2.5 kpc)$^{2}$ L$_{\odot}$, and a quiescent (V$\sim$10.5) accretion disc bolometric luminosity of about 300 $(d/2.5 kpc)^2$ L$_{\odot}$. These correspond to maximum and quiescent accretion rates of 6 $\times 10^{-7}$ and 1 $\times~10^{-7}$ (d/2.5 kpc)$^2$ (R / 0.01 R$_{\odot}$) (0.9 M$_{\odot}$/ M)~M$_{\odot}$ yr$^{-1}$, respectively (assuming $\dot{M}$=LR/GM). The reddening, E(B-V)=0.15 $\pm$ 0.05 (\citealt{Schmid2001}, and our Appendix~\ref{app1}), introduces a $\pm$25\% uncertainty in the accretion rates and luminosities, in addition to any uncertainty in the distance or in disentangling the RG flux from the accretion disc. In Appendix~\ref{app1}, we review and update the \citet{Meier1996} and \citet{Schmid2001} case for a distance to MWC 560 of 2.5 kpc, including new supporting evidence. In Appendix~\ref{app7}, we report our method for calculating the accretion rates and luminosities.

The estimated 2012--2014 quiescent accretion rate is very high for a symbiotic star with flickering (i.e., without WD surface burning). Comparison to some theoretical expectations for WD surface burning as a function of WD mass and accretion rate, as in figure 1 of \citet{Wolf2013}, yields interesting results. For WD masses less than 0.9 M$_{\odot}$, MWC 560's quiescent accretion rate would be expected to lead to stable burning, inconsistent with observations---but the large hard X-ray component temperature observed in MWC 560 by \citet{Stute2009} does imply a high white dwarf mass, so this is not a problem. At 0.9 M$_{\odot}$, stable burning is narrowly avoided, but a recurrent nova is expected every century or so; in other words, MWC 560 may be due for an imminent nova. For WD masses higher than 0.9 M$_{\odot}$, the nova recurrence time quickly becomes too short, inconsistent MWC 560's 9-decade-long optical light curve \citep{Leibowitz2015}. If the WD mass is higher than 0.9 M$_{\odot}$ (as perhaps suggested by the comparable boundary layer cooling flow temperatures between MWC 560 and RT Cru / T CrB; \citealt{Stute2009}), or a nova does not occur sometime soon, that may support competing theoretical pictures (e.g., \citealt{Starrfield2012a,Starrfield2012b}) or suggest an important role for outflows in preventing novae. However, some caution is warranted due to the uncertainties introduced by flux modelling, reddening, and especially distance. A more careful assessment of MWC 560's possibly liminal location in WD mass / accretion rate parameter space may be possible after future GAIA data releases with binary solutions and smaller uncertainties.

\subsection{Self-regulation of the disc by its outflow}\label{4.3}

Differences between brightening events in 1990 and 2016 suggest that the 1-month outflow burst in January--February 2015 and the sustained outflow fast state in January--July 2016 (\autoref{fig2}) may have helped stabilize the accretion system and helped keep the inner disc intact during an increase in accretion rate through the disc. Dramatic disruptions in the outflow and the disc in 1990 did not repeat in 2016, even though both events reached similar peak accretion rates.

\subsubsection{Historical overview}

In the 9 decades of observations preceding 2016 \citep{Luthardt1991,Leibowitz2015,Munari2016}, two other optical brightening events came close to the maximum optical luminosity of 2016. The 2010-2011 event reached at least V$\approx$9.6, and was accompanied by absorption velocities as high as -4200 km s$^{-1}$ \citep{Goranskij2011}. The 1989--1990 event reached at least V$\approx$9.2, and was accompanied by rapidly variable broad absorption lines with velocities as fast as -6000 km s$^{-1}$ \citep{Tomov1990,Tomov1992}. 1989--1990 may have included a step-function-like shift in the light curve, separate from the postulated orbital and RG periodicities that may explain the timing of the three events \citep{Leibowitz2015}. At least, 1990 is consistent with a secular brightening over the last century \citep{Munari2016}; the three optically-brightest events have all occurred during or after 1990, and the average optical brightness after 1990 has been about triple the average optical brightness before 1990 \citep{Leibowitz2015}.

\subsubsection{1990 vs. 2016}

In one respect, 1990 and 2016 were similar: using the same distance and reddening, and a similar flux model, \citet{Schmid2001} obtained about the same peak accretion rate and luminosity for the 1990 event as we did for 2016 (\autoref{4.2}). But otherwise, the 1990 and 2016 events differed starkly.

{\it (1)} In 1990, the Balmer absorbers were rapidly variable, appearing and vanishing in matters of days, detached from their associated emission lines, and ejected at velocities up to -6000 km s$^{-1}$ \citep{Tomov1990,Tomov1992}; then, later in 1990, the outflow slowed to a very unusually small $\lesssim 900$ km s$^{-1}$ until about a year after the outburst \citep{Zamanov2011a}. In 2016, the outflow was much more stable (\autoref{opticalabs}: Balmer lines varied only slightly (slowly and predictably in time with the optical/NUV flux, likely a photoionization effect; Appendix~\ref{app3}), never vanished, stayed contiguous with their emission, and never went faster than $\approx3000$km s$^{-1}$; then, later in 2016, the Balmer lines slowed, but never fell below $\approx1500$km s$^{-1}$.

{\it (2)} Over the course of the 1990 outburst, the veil of \ion{Fe}{ii} absorption in the UV lifted, clearing entirely by the end of April 1990 \citep{Lucy2018,Maran1991}; as discussed in \citet{Lucy2018}, this likely represented a temporary switch from persistent outflow to discrete mass ejection, resembling in the latter state the P Cygni phase of some novae. In 2016, this iron curtain remained optically thick throughout and the spectral morphology did not vary significantly (\autoref{uvspectrasec}), consistent with the stable and persistent outflow that we saw in the optical.

{\it (3)} Fast optical flickering was partially suppressed during the 1990 outburst, and thoroughly suppressed by the next observing season later in that year \citep{Zamanov2011a}. U band fast photometry collected during the 1990 outburst peak and discussed in \citet{Zamanov2011a} show comparatively smooth light curves. In 6 hours of U band photometry by T. Tomov over 4 days during the 1990 high state, provided to us by R. Zamanov (2017, private communication), there were slow 0.1 mag oscillations on hour time-scales, and on shorter time-scales one incident of 0.2 mag flickering and a few incidents of 0.1 flickering. Afterwards, at the same time as the Balmer velocities were unusually slow, essentially no flickering at all was observed on time-scales of minutes \citep{Zamanov2011a}. In contrast, flickering persisted throughout 2016 (\autoref{flickering}); there was no evidence at all for suppression (though poor sampling after mid-2016 warrants a little caution).

In brief, it appears that in 1990, MWC 560 underwent a dramatic disruption of the flicker-producing inner disc \citep{Zamanov2011a}, perhaps as a result of ejections of mass evacuating the inner disc, and that this disruption interrupted the launching mechanism for the high-velocity outflow until the inner disc was rebuilt a year later.  While no dramatic colour changes were observed in 1990, reprocessing of the accretion disc light by an optically thick, low velocity wind, like that proposed for MWC 560 by \citet{Panferov1997}, could make any changes in accretion disc size and temperature profile difficult to observe in the UV and optical. Disc evacuation has also been proposed to have occurred in the symbiotic star CH Cyg, which is believed to have once been evacuated by a jet, causing its flickering to temporarily cease \citep{Sokoloski2003a,Sokoloski2003b}. Such incidents also have precedent in X-ray binaries, as recently demonstrated by V404 Cyg \citep{Munoz-Darias2016}. \citet{Neilsen2011} also suggested that accretion discs in X-ray binary systems like GRS 1915+105 may suppress themselves by carrying away the mass of the inner-disc that launches them. 

\begin{figure*}
\centering{\includegraphics[width=7in]{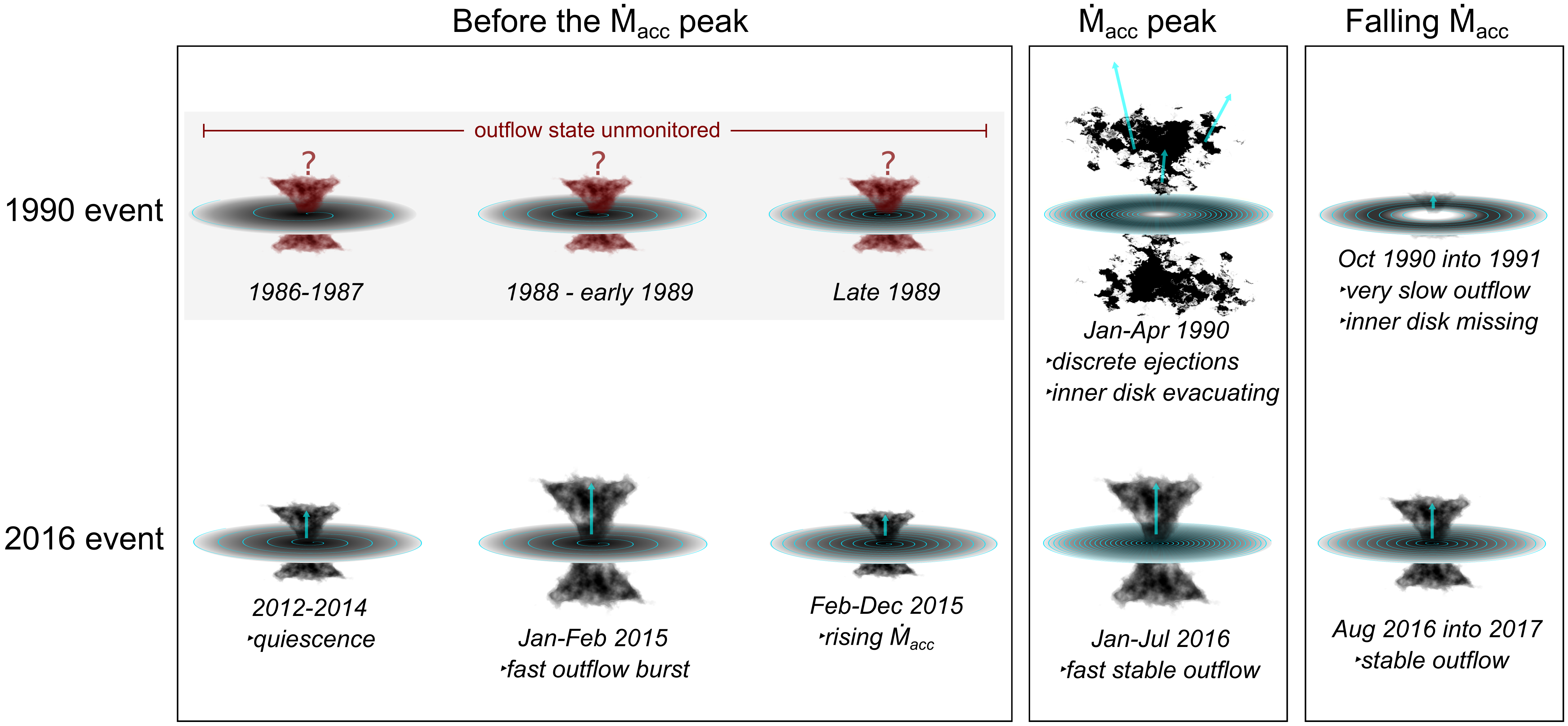}}
\caption{We propose that the accretion disc outflow in MWC 560 regulates the disc, facilitating an evacuation of the inner disc in 1990 through catastrophic discrete mass ejections, and inhibiting this process 26 years later through a steady fast outflow at the 2016 accretion rate peak and through an outflow burst just before the accretion rate started to rise in 2015. The 1990 (top row) and 2016 (bottom row) events differed starkly. The density of the blue spirals on the discs signifies the accretion rate through the disc; more spiral turns indicates a higher accretion rate. The length of the blue arrows and the spatial extent of the outflow signify the maximum outflow velocity; the state of the outflow from 1986 through 1989 is unknown, so those phases are marked with question marks and red placeholder outflows. The evacuation of the inner disc in 1990 is denoted with holes in the centre of the discs/spirals. {\it The opening angles, scale, and morphology of the outflow in these drawings are arbitrary; as discussed in \citet{Lucy2018}, the geometry of the outflow is not well constrained.}} \label{fignewnew}
\end{figure*}

In 2016, none of that happened. The outflow power increased abruptly as the system reached its peak accretion rate through the visual-emitting disc, but the new outflow was stable, the inner disc was intact, and neither the outflow nor the inner disc were destroyed by the event. \autoref{fignewnew} shows a schematic diagram of MWC 560 before, during, and after the 1990 and 2016 accretion rate peaks, illustrating the difference between these two events.

\subsubsection{The outflow as a gatekeeper}

Two interrelated processes may explain the stability of the disc and the outflow in 2016: the timely evacuation in the outflow of mass from the accretion disc, and a long-term trend towards higher average accretion rates over the course of the last century. We propose that the outflow helped maintain equilibrium in the disc, slowing down changes in the disc in 2015, finally halting changes in the disc in 2016, and thereby preventing the inner disc's evacuation.

There was a month-long expulsion of high-velocity material in January--February 2015 (\autoref{fig2}), just before the optical flux began to slowly rise. Perhaps an early-2015 disc instability or an increase in the RG mass transfer rate, which would not necessarily immediately be detectable in the optical depending on where in the disc it manifested, was slowed and stabilized by a wind burst carrying away excess mass and angular momentum. Although optical rises reliably predict increased outflow velocity throughout the history of MWC 560 \citep{Tomov1990,Iijima2001,Goranskij2011} in a clear cause-and-effect relationship, such outflow velocity jumps have previously occurred in optical flux quiescence too \citep{Iijima2001}; the system clearly varies in ways that we sometimes cannot detect, and the outflow could play a role in suppressing the propagation of these changes through the disc.

 The outflow might suppress any heating or cooling wave that reaches its launching radius in the disc. After February 2015, the accretion rate through the visual-emitting disc continued to rise (\autoref{fig2}), but the outflow remained slow and static until January 2016, when it jumped in power and subsequently persisted steadily for months as the optical light curve flattened (\autoref{4.1}). We speculate that the delay between the start of the optical rise and the initiation of a persistent fast outflow could be due to feedback from the outflow burst and the multiple directions in which heating and cooling waves can travel through an accretion disc. The propagation of heating and cooling waves across the disc in a dwarf-nova-type outburst can involve mass being transported along the angular momentum gradient of the disc in both directions \citep[e.g.,][]{Lasota2001,Schreiber2003}. It may be that the fast outflow launching radius was roughly the last radius to experience the outburst responsible for the 2016 accretion rate peak. For example, an outside-in outburst might only affect a centrally-launched wind after the increased accretion rate finally propagated to the centre of the disc, at which point a fast outflow could launch while the disc stabilized. The 2016 optical rise occurred near the confluence of two long-term periodicities in MWC 560's light curve \citep{Leibowitz2015}. A rise in the external accretion rate (i.e., rate of mass flow into the disc) as periastron approached during a period of high RG mass loss could induce multiple instabilities in the disc; those at the outflow launching radius would be suppressed by an outflow burst \citep[and by the difficulty of propagating to higher surface densities and angular momenta; e.g.,][]{Schreiber2003}, while others would have time to propagate through the disc before restarting the outflow, themselves getting suppressed by the 2016 outflow fast state.
 
  What physical differences between 2016 and 1990 allowed the formation of a stable fast outflow in 2016, instead of driving catastrophic discrete evacuations of the inner disc as in 1990? The optical light curve in 2014--2016 was about the same as in 1988--1990, but the 2006--2013 period was typically at least twice as bright as the 1980--1987 period \citep{Munari2016,Doroshenko1993,Luthardt1991}, consistent with a secular brightening throughout the last century \citep{Munari2016} or with a step-function tripling of the average optical luminosity in 1989--1990 \citep{Leibowitz2015}. So it may be that the structure of the disc (and therefore its outflow-driving mechanism) was acclimatized, on a time-scale of at least several years, to a much higher accretion rate through the disc in 2016 than in 1990. The large discs of symbiotic stars plausibly allow for viscous time-scales of at least several years to govern the nature of the as-yet-undetermined outflow-driving mechanism. Alternatively, the difference between 2016 and 1990 could involve changes in the boundary layer that are not always traced by the optical flux; the disappearance of hard X-rays between 2007 and 2013, and their continued absence in our 2016 observations, might indicate that the optical depth of the boundary layer had been increasing over at least the last decade. Future observations could test whether an optically thick boundary layer is necessary to initiate a stable outflow without evacuating the inner disc in discrete mass ejections.

\citet{Neilsen2011} suggests that, in the X-ray binary microquasar GRS 1915+105, a high-mass (\.{M}$_{wind}$ $\geq$ 15\.{M}$_{acc}$) wind may be ``effectively acting as a gatekeeper or a valve for the external accretion rate, and facilitating or inhibiting state transitions'' (in that case, via \citealt{Shields1986} oscillations). Our 2016 observations provide strong new evidence that the role of MWC 560's outflow is similar to the role of GRS 1915+105's outflow, a comparison first made in broader terms by \citet{Zamanov2011a}. During the 2016 outflow fast state of MWC 560, the mass outflow rate estimated from our radio observations (perhaps $\gtrsim10^{-6}$~(d/2.5 kpc)$^{5/2}$~M$_{\odot}$~yr$^{-1}$ {\it if} the outflow has uniform density; \autoref{4.1.2}) may have been commensurate with the accretion rate through the disc (6 $\times$ 10$^{-7}$ (d/2.5 kpc)$^{2}$ (R / 0.01 R$_{\odot}$) (0.9 M$_{\odot}$/ M) M$_{\odot}$ yr$^{-1}$; \autoref{4.2}, Appendix~\ref{app7}), although neither is sufficiently well constrained for this comparison to be more than a consistency check. The possible evacuation of the inner disc of MWC 560 during the 1990 outburst \citep{Zamanov2011a}, accompanied by the phenomenology of discrete mass ejection in 1990 \citep{Lucy2018}, suggests that MWC 560's outflow may indeed be capable of acting as a valve, facilitating an accretion disc state transition in 1990 and inhibiting it in 2016.

\section{Conclusions} \label{conclusions}

If accretion disc outflows are common in symbiotic stars \citep[e.g.,][]{Lucy2018,Sokoloski2003c,Muerset1997}, then the effect of outflows on their accretion discs could be fundamental to understanding their physics and evolution. The broad absorption line symbiotic star MWC 560 = V694 Mon is an important laboratory for this complex relationship. We conducted coordinated radio, optical, NUV, and X-ray spectroscopic and photometric observations of MWC 560 during an optical flux maximum in 2016, and we detected an outflow fast state from January 2016 into likely the end of July 2016. The variability in each band is plotted in \autoref{fig2} for the years 2012 through 2016.

\begin{enumerate}[leftmargin=*,rightmargin=0ex, label=(\arabic*)]
\itemsep0.3em 

\item The maximum velocity of a dense, Balmer line-absorbing outflow from MWC 560's accretion disc abruptly doubled over the course of at most 3 weeks in January 2016, near the 2016 February 7 accretion rate peak and at the completion of a year-long rise in the accretion rate through the visual-emitting disc. The abrupt change in Balmer absorption velocity was almost certainly not due to photoionization effects, although a subsequent correlation on week time-scales between high-velocity Balmer opacity and optical/NUV flux can probably be attributed to photoionization. High velocities were stable and sustained through at least mid-April of the same year, before beginning a slow decline; even through to the end of 2016, velocities in high-resolution spectra never dropped below 1500~km~s${^-1}$. The density of the Balmer-absorbing gas was $\gtrsim10^{6.5}$~cm$^{-3}$. (\autoref{opticalabs}, \autoref{4.1.1})

\item Radio emissions confirm an increase in outflow power at the onset of higher Balmer absorption velocities. Flat-spectrum radio emissions detected at 3.1, 9.8, and 33.1 GHz began to rise linearly at a rate of about 20 $\mu$Jy/month, even as the optical/NUV flux varied up and down. The slope of this radio rise suggests that it started at its 37 $\mu$Jy quiescent value (observed in 2014) in January 2016, around the same time as the high-velocity Balmer absorption appeared. Radio emissions reached a maximum of 175 $\pm$ 10 $\mu$Jy at 9.8 GHz on 2016 July 29 before beginning a slower decline. The emission mechanism was thermal and optically-thin, originating in gas with a density $\lesssim10^{5.5}$~cm$^{-3}$. We propose that this lower-density region was steadily fed by the denser Balmer absorption-line fast outflow. (\autoref{resultsradio}, \autoref{4.1.2})

\item Also during the 2016 outflow fast state, soft X-rays were observed to be brighter by an order of magnitude relative to both prior X-ray epochs (2013 and 2007). The plasma temperature of this component, constrained to between kT $\approx$ 0.1 and 1 keV, was consistent with strong-shock velocities of 300-900 km s$^{-1}$, in turn consistent with differential velocities in the optical Balmer absorption lines. Meanwhile, the hard X-ray component in both 2016 and 2013 was about a third as bright as it had been in 2007. (\autoref{resultsxrays}, \autoref{4.1.3}, \autoref{vs2013})

\item Balmer velocity profiles observed close to each of the three X-ray epochs suggest that soft X-rays are weak when only high-velocity absorption is present (2007) and when only low-velocity absorption is present (2013); only in 2016, when both high-velocity and low-velocity Balmer absorption was observed, was the soft X-ray flux strong. We propose that the soft X-ray component originates in a shock where these new fast absorbers and pre-existing slow absorbers in the absorption-line outflow collide. (\autoref{4.1.3}, \autoref{fignew}).

\item The iron curtain of overlapping absorption in the NUV remained optically thick throughout the 2016 outflow fast state, without substantial variability in its spectral morphology, further supporting the presence of a stable and sustained outflow. (\autoref{uvspectrasec}, \autoref{4.1.1})

\item Optical/NUV flickering shows that the inner accretion disc remained intact in 2016; the slow optical brightening from 2015 into 2016, which led to the outflow power jump, was due to an increase in the rate of accretion through the disc. Flickering by at least a typical 0.1 mag per 15--30 minutes in V band, up to 0.4 magnitudes over 10 minutes in the NUV, persisted throughout the 2016 outflow fast state---similar to that observed in quiescence and high states from the end of 1991 through 2015. This flickering requires that MWC 560's luminosity be powered by accretion alone without WD surface burning. The high accretion rate, even in the relative quiescence of 2012--2014, may suggest that a nova should be expected in MWC 560 within the next century. (\autoref{flickering}, \autoref{4.2})

\item The peak accretion rate in 2016 was about the same as the peak accretion rate in the 1990 outburst of the same system. Despite this, the two events differed dramatically: In 1990, rapidly variable mass ejections up to 6000 km~s$^{-1}$ appeared to evacuate the inner accretion disc, leading to the cessation of flickering and the suppression of the outflow to velocities below 900 km~s$^{-1}$ for up to a year, and the temporary disappearance of iron curtain absorption in the UV. Throughout 2016, the outflow was stable and the inner disc remained intact; strong and rapid flickering continued, Balmer velocities reached up to 3000 km~s$^{-1}$ and slightly varied in time with the optical/NUV flux, and the iron curtain remained optically thick. (\autoref{4.3})

\item We propose that the outflow sometimes inhibits structural changes in the accretion disc and sometimes facilitates them. In 1989--1990, the outflow evacuated the inner disc in a phase of discrete mass ejection. But in 2015--2016, a 1-month outflow velocity burst in January--February 2015 and the longer, stable 2016 outflow fast state may have prevented a catastrophic evacuation of the inner disc, by carrying away excess accreting mass and suppressing heating and cooling waves as they reached the outflow's launching radius. The difference between the disc/outflow relationship in 2016 versus 1990, which reached similar peak accretion rates, might be due to a secular increase in the decade-averaged accretion rate throughout the last century. The complex, self-regulatory relationship between a symbiotic star accretion disc and its outflow resembles X-ray binary behaviour. (\autoref{4.3}, \autoref{fignewnew})

\end{enumerate}

\section*{Acknowledgements}

With thanks to Neil Gehrels. 

We thank Fred Hamann for helping us understand the relationship between Balmer absorption and photoionization, Elena Barsukova and Vitaly Goranskij for generously sending additional spectra, and Josh Peek and Kirill Tchernyshyov for assistance with their ISM velocity map. We acknowledge Steve Shore for wisely encouraging ARAS observations in early 2015.

We thank volunteer observers Teofilo Arranz, Gary Walker, and Geoffrey Stone, whose data were submitted to the AAVSO and are reproduced in \autoref{vflick}. We are similarly indebted to all the other AAVSO, ARAS, and independent observers across the world whose work made this paper possible: H. Adler, S. Aguirre, T. Atwood, D. Barrett, P. Berardi, B. Billiaert, D. Blane, E. Blown, J. Bortle, D. Boyd, J. Briol, C. Buil, F. Campos, A. Capetillo Blanco, R. Carstens, J. Castellani, W. Clark, T. Colombo, A. Debackere, X. Domingo Martinez, S. Dvorak, J. Edlin, J. Foster, R. Fournier, L. Franco, J. Garlitz, A. Garofide, A. Glez-Herrera, K. Graham, C. Gualdoni, J. Guarro Flo, J. Foster, F. Guenther, C. Hadhazi, F. Hambsch, B. Harris, D. Jakubek, P. Lake, T. Lester, S. Lowther, C. Maloney, H. Matsuyama, K. Menzies, V. Mihai, J. Montier, G. Murawski, E. Muyllaert, G. Myers, P. Nelson, M. Nicholas, O. Nickel, S. O'Connor, J. O'Neill, W. Parentals, A. Plummer, G. Poyner, S. Richard, J. Ripero Osorio, J. Ritzel, D. Rodriguez Perez, R. Sabo, L. Shotter, P. Steffey, W. Strickland, D. Suessman, F. Teyssier, T. Vale, P. Vedrenne, A. Wargin, and  A. Wilson. Many of these observers' data were reproduced in \autoref{fig2}. We apologize to any observers we neglected to acknowledge, with deep gratitude. We thank the staff of the AAVSO International Database and the ARAS Spectral Data Base, including E. Waagen and F. Teyssier.

We thank the Chandra team (PI: B. Wilkes), the Swift team (PI: N. Gehrels), and the Very Large Array team for the discretionary time, and for their technical help throughout the observations on which this paper is based. ABL thanks K. Mukai, T. Nelson, T. Iijima, the Chandra help desk, the Swift help desk, and the GAIA help desk for useful conversations.

ABL is supported by the NSF GRFP under grant DGE-1644869. ABL and JLS are supported by Chandra award DD6-17080X. JLS is supported by NSF AST-1616646. UM is partially supported by PRIN INAF 2017 (Towards the SKA and CTA era: discovery, localisation and physics of transient sources, P.I. M. Giroletti). NR acknowledges support from the Infosys Foundation through the Infosys Young Investigator grant. GJML is a member of the CIC-CONICET (Argentina) and acknowledges support from grant \#D4598, ANPCYT-PICT 0478/2014 and 0901/2017. MJD acknowledges financial support from the UK Science and Technology Facilities Council. This work was supported in part by the UK Space Agency. ABL thanks the LSSTC Data Science Fellowship Program, which is funded by LSSTC, NSF Cybertraining Grant \#1829740, the Brinson Foundation, and the Moore Foundation; their participation in the program has benefited this work.

We gratefully acknowledge our use of pysynphot \citep{pysynphot}, PyAstronomy (\url{https://github.com/sczesla/PyAstronomy}), extinction \citep{extinction}, the Mikulski Archive for Space Telescopes (MAST), the International Ultraviolet Explorer (IUE), the Sloan Digital Sky Survey (SDSS), NASA's Astrophysics Data System (ADS), the SIMBAD database operated at CDS, and the National Institute of Standards and Technology (NIST).

Some of the data presented in this paper were obtained from the Mikulski Archive for Space Telescopes (MAST). STScI is operated by the Association of Universities for Research in Astronomy, Inc., under NASA contract NAS5-26555. Support for MAST for non-HST data is provided by the NASA Office of Space Science via grant NNX09AF08G and by other grants and contracts. The Liverpool Telescope is operated on the island of La Palma by Liverpool John Moores University in the Spanish Observatorio del Roque de los Muchachos of the Instituto de Astrofisica de Canarias with financial support from the UK STFC. This work was based in part on data obtained with the Asiago 1.82m Copernico (INAF Padova) and Asiago 1.22m Galileo (University of Padova) telescopes.




\bibliographystyle{mnras}
\bibliography{lucy} 



\appendix

\section{Distance, the burning question} \label{app1}

In this section, we review and update the case for a distance of about 2.5 kpc \citep[][]{Schmid2001,Meier1996}, which yields high accretion rate estimates that may be in some tension with some empirical and theoretical expectations for typical non-burning symbiotics, or suggest that a thermonuclear nova is imminent (\autoref{4.2}). Each individual argument relies on debatable assumptions and the uncertainty is poorly constrained, but d$\approx$2.5 kpc is supported by several largely-independent lines of reasoning.

{\it Geometric parallax:} Adopting the \citet{Bailer2018} Bayesian analysis of GAIA DR2, the parallax distance is 2.4$^{+1.3}_{-0.7}$ kpc. Here we assume that the parallax is not affected by binary motion, and that the Bayesian prior employed works for red giants in general and symbiotic stars in particular.

{\it Reddening and dust maps:} 
Three pieces of evidence point to an E(B-V) between 0.1 and 0.2: fits of field giant templates to the optical spectrum, the 2200\AA\ absorption bump \citep{Schmid2001}, and the strength of interstellar absorption lines. For our part, we performed a fit of 5 Gaussians to the \ion{Na}{i}D complex (one Gaussian for each of 4 resolved absorption lines, and one for what we suspect is remnant sky emission) in the publicly available pipeline reduction of the FEROS 07881 1990 November 14 spectrum, following the method for resolved lines described by \citet{Munari1997}. We obtained a total reddening of E(B-V)=0.1, which we regard as a lower limit due to uncertainties in the continuum placement and underlying emission. Using the Bayestar2017 3D map of Galactic extinction by \citet{Green2018} and incorporating their reported statistical uncertainties, E(B-V)$\geq$0.1 corresponds to a distance $\geq$2.1 kpc. The {\it Stilism} 3D dust map v. 4.1 is even more consistent with our conclusions, yielding E(B-V)=0.15 at 2.5 kpc \citep{Lallement2018,Capitanio2017}. However, the residuals between all types of 3D extinction maps and individual stars are often significant (e.g., see figure 9 in \citealt{Green2015}, and severe angular smoothing in \citealt{Lallement2018}).

{\it Optical flux:} For E(B-V)=0.15, the best-fitting RG in the optical (Appendix~\ref{app7}) has a V band apparent magnitude of 11.8, corresponding to a 2.5 kpc distance. Here we assume that we have correctly disentangled the giant's optical flux from the accretion disc flux, and that the giant has an absolute magnitude M$_{v}$ = -0.4 -- -0.5 consistent with both bulge and field giants \citep{Allen1983,Whitelock1992}.

{\it Infared flux:} MWC 560 has an average J band magnitude of 6.5, and \citet{Meier1996} used this to obtain a distance of 2.5 kpc. For a field giant with (V-J)$_{0}$=5.4 \citep{Koornneef1983} and E(B-V)=0.15, corresponding to A$_{j}$=0.87$\times$E(B-V)=0.13 \citep{Savage1979}, we obtain a distance of 2.9 kpc. Similar results are obtained in the K band. If we assume that the accretion disc and the giant have the same reddening, the observed (V-J)$_0$ of MWC 560 is more consistent with a field giant than a bulge giant. Still, if we did instead assume a bulge giant with higher reddening than the accretion disc, we would obtain a distance of about 1 kpc, illustrating the large model uncertainties in flux-based distance estimation.

{\it Interstellar absorption line velocities:} \citet{Schmid2001} argued for a large distance based on the radial velocities of apparently interstellar \ion{Na}{i}D and \ion{Ca}{ii} absorption. In the LSR frame, the resolved lines they observed actually correspond to -7, 0, +27, +35, and +52 km s$^{-1}$. LSR radial velocities in the Galactic disc are typically around +10 km s$^{-1}$ above a flat rotation curve (V$_{\odot}$=220 km/s, d$_{\odot}$=8.5 kpc) at this longitude, yielding +23 and +39 km s$^{-1}$ at 1 and 2.5 kpc, respectively \citep[][and private communication]{Tchernyshyov2017}. This indeed supports a large distance $\gtrsim$2.5 kpc, although the affects of turbulence in the ISM are poorly constrained \citep[e.g.,][]{Heiles2003}.

\section{Optical and UV observation metadata}
\label{app2}

({\it appendices continue on following pages})

\begin{table*}
\begin{threeparttable}
\begin{tabular}{lccl}
\hline
Start time (UT) & Exposure time (s) & Roll angle ($deg$) & Quality flags\tnote{*}\\
\hline
2016 Mar 02 07:47 & 785.2 & 251 & ZOc:2300-2376,2590-2650 \\
2016 Mar 05 12:09 & 892.5 & 254 & ZOc:2790-2890,2880-3200 \\
2016 Mar 09 12:00 & 307.8 & 251 & ZOc:2300,2376,2590-2650 \\
2016 Mar 12 08:39 & 463.3 & 254 & ZOc:2790-2890,2880-3200” \\
2016 Mar 17 22:25 & 892.5 & 256 & ZOc:2285-2385 \\
2016 Mar 20 22:19 & 999.8 & 260 & ZOc:2720-2820 \\
2016 Mar 23 22:03 & 470.1 & 260 & ZOc:2740-2810 \\
2016 Mar 26 17:00 & 1008.6 & 264 & ZOc:$<$1886 \\
2016 Mar 29 10:25 & 946.6 & 268 & ZOc:$<$1775,1905-2005,3230-3330; bright FO nearby \\
2016 Apr 01 15:00 & 892.5 & 264 & ZOc:$<$1870,FOc:$>$3700 \\
2016 Apr 04 14:49 & 892.5 & 268 & ZOc:$<$1720,1930-2000,2180-2255,2300-2417,2547-2600,3240-3325 \\
2016 Apr 20 09:03 & 892.5 & 281 & ZOc:$<$1756,2400-2440w,2540-2600,2935-3110 \\
2016 Apr 27 05:06 & 892.5 & 283 & ZOc:1880-1947,2006-2090,2510-2640 \\
2016 May 04 14:16 & 463.3 & 284 & ZOc:2012-2070,2525-2622,3025-3127; bright FO nearby \\
2016 May 11 04:02 & 892.5 & 286 & ZOc:1970-2010,2250-2300; fluxes underestimated: loss of lock \\
2016 May 18 17:54 & 892.5 & 297 & ZOc:2620-2720 \\
2016 Jun 01 23:17 & 845.9 & 307 & ZOc:$<$1990 \\
\hline
\end{tabular}
\begin{tablenotes}
\caption{Swift UV spectra. Observation metadata and quality flags for Swift UVOT UV Grism spectra. \label{tableuv}}
\item[*] \footnotesize Manually determined quality flags include ZOc = zeroth order contamination at the indicated wavelengths (\AA), FOc = first order contamination at the indicated wavelengths (\AA), FO = first order, and individualized whole-spectrum flags.
\end{tablenotes}
\end{threeparttable}
\end{table*}

\begin{table}
\caption{Optical spectra. ECH = Echelle spectrograph on the Asiago 0.61m or 1.82m telescopes (R=18000-20000). 
B\&C = B\&C spectrograph on Asiago 1.22m telescope (R=1300).
LT = Liverpool Telescope (R$\sim$2500). 
UCS8 = home-built spectrograph in L'Aquila (R$\sim$1300).
PSO = Shelyak LHires III spectrograph in Tata (resolution varies and is specified in table). For more details, see \autoref{obsoptical}. \label{tablespectra}}
\resizebox{0.44 \textwidth}{!}{
\begin{tabular}{lccc}
\hline
Date & Time (UT) & Exposure (s) & Telescope+Instrument\\
\hline
2007 Oct 30 & 03:01 & 4500 & 0.61m+ECH \\
2012 Jan 27 & 21:40 & 900 & 1.22m+B\&C \\
2012 Feb 08 & 22:10 & 600 & 1.82m+ECH \\
2012 Mar 31 & 18:52 & 480 & 1.22m+B\&C \\
2013 Jan 26 & 22:13 & 900 & 1.22m+B\&C \\
2014 Feb 09 & 21:22 & 480 & 1.22m+B\&C \\
2014 Mar 14 & 20:08 & 600 & 1.82m+ECH \\
2014 Oct 30 & 02:30 & 900 & 1.22+B\&C \\
2015 Mar 08 & 20:24 & 900 & 1.82m+ECH \\
2015 Mar 10 & 19:23 & 600 & 1.22m+B\&C \\
2015 Oct 02 & 02:56  & 3620 & PSO (R$\sim$450) \\
2015 Oct 04 & 02:41 & 4433 & PSO (R$\sim$14000) \\
2015 Dec 31 & 00:37 & 4959 & PSO (R$\sim$3300) \\ 
2016 Jan 21 & 21:23 & 300 & 1.22m+B\&C \\
2016 Feb 05 & 17:49 & 4500 & 0.61m+ECH \\
2016 Feb 05 & 19:07 & 480 & 1.22m+B\&C \\
2016 Feb 05 & 22:55 & 2800 & USC8 \\
2016 Feb 11 & 19:44 & 2800 & USC8 \\
2016 Feb 20 & 20:20 & 900 & 1.82m+ECH \\
2016 Feb 20 & 21:08 & 480 & 1.22m+B\&C \\
2016 Feb 20 & 17:58 & 2800 & USC8 \\
2016 Feb 23 & 18:25 & 3600 & 0.61m+ECH \\
2016 Feb 25 & 18:32 & 3600 & 0.61m+ECH \\
2016 Mar 01 & 17:58 & 3600 & 0.61m+ECH \\
2016 Mar 02 & 19:52 & 3600 & USC8 \\
2016 Mar 03 & 18:54 & 4500 & 0.61m+ECH \\
2016 Mar 04 & 18:38 & 3200 & USC8 \\
2016 Mar 06 & 20:31 & 4500 & 0.61m+ECH \\
2016 Mar 06 & 18:10 & 3200 & USC8 \\
2016 Mar 07 & 18:15 & 3600 & 0.61m+ECH \\
2016 Mar 07 & 20:14 & 60 & LT \\
2016 Mar 07 & 23:46 & 60 & LT \\
2016 Mar 08 & 18:22 & 4500 & 0.61m+ECH \\
2016 Mar 08 & 20:25 & 3600 & 0.61m+ECH \\
\end{tabular}
}
\end{table}

\begin{table}
\renewcommand\thetable{B2} 
\caption{-- {\it continued}}
\resizebox{0.35 \textwidth}{!}{
\begin{tabular}{lccc}
2016 Mar 08 & 22:40 & 60 & LT \\
2016 Mar 08 & 00:03 & 60 & LT \\
2016 Mar 08 & 20:11 & 60 & LT \\
2016 Mar 08 & 21:22 & 60 & LT \\
2016 Mar 08 & 18:23 & 3600 & USC8 \\
2016 Mar 09 & 20:14 & 60 & LT \\
2016 Mar 09 & 22:58 & 60 & LT \\
2016 Mar 10 & 18:34 & 4500 & 0.61m+ECH \\
2016 Mar 10 & 18:15 & 480 & 1.22m+B\&C \\
2016 Mar 11 & 18:43 & 5400 & USC8 \\
2016 Mar 13 & 18:13 & 360 & 1.22m+B\&C \\
2016 Mar 14 & 21:32 & 1800 & 0.61m+ECH \\
2016 Mar 17 & 19:01 & 5400 & 0.61m+ECH \\
2016 Mar 18 & 19:03 & 5400 & 0.61m+ECH \\
2016 Mar 18 & 18:40 & 5400 & USC8 \\
2016 Mar 19 & 19:08 & 900 & 1.82m+ECH \\
2016 Mar 19 & 18:58 & 360 & 1.22m+B\&C \\
2016 Mar 19 & 18:53 & 4800 & USC8 \\
2016 Mar 21 & 18:58 & 5400 & 0.61m+ECH \\
2016 Mar 23 & 18:42 & 4500 & 0.61m+ECH \\
2016 Mar 24 & 18:34 & 3600 & 0.61m+ECH \\
2016 Apr 03 & 19:13 & 300 & 1.22m+B\&C \\
2016 Apr 06 & 19:10 & 3600 & 0.61m+ECH \\
2016 Apr 10 & 18:53 & 3600 & 0.61m+ECH \\
2016 Apr 12 & 18:54 & 480 & 1.22m+B\&C \\
2016 Apr 14 & 19:39 & 3600 & 0.61m+ECH \\
2016 Apr 19 & 19:09 & 3600 & 0.61m+ECH \\
2016 Apr 19 & 19:10 & 360 & 1.22m+B\&C \\
2016 Apr 27 & 19:24 & 3600 & 0.61m+ECH \\
2016 May 06 & 19:27 & 300 & 1.22m+B\&C \\
2016 Oct 05 & 03:07 & 4500 & 0.61m+ECH \\
2016 Oct 13 & 03:41 & 900 & 1.82m+ECH \\
2016 Oct 23 & 03:22 & 1200 & PSO (R$\sim$ 2500) \\ 
2016 Dec 04 & 01:15 & 600 & PSO (R$\sim$450) \\
2016 Dec 11 & 01:38 & 3600 & PSO (R$\sim$450) \\   
2016 Dec 16 & 01:48 & 1200 & 1.82m+ECH \\
2017 Jan 08 & 20:43 & 600 & PSO (R$\sim$450) \\
2017 Jan 20 & 20:43 & 1800 & PSO (R$\sim$750) \\
2017 Jan 21 & 20:43 & 2400 & PSO (R$\sim$750) \\
2017 Jan 22 & 20:43 & 600 & PSO (R$\sim$750)\\
\hline
\end{tabular}
}
\end{table}

\newpage
\newpage

~

~

\section{Absorption correlated with flux: a photoionization effect?} \label{app3}

\begin{figure*}
\centering{\includegraphics[width=7in]{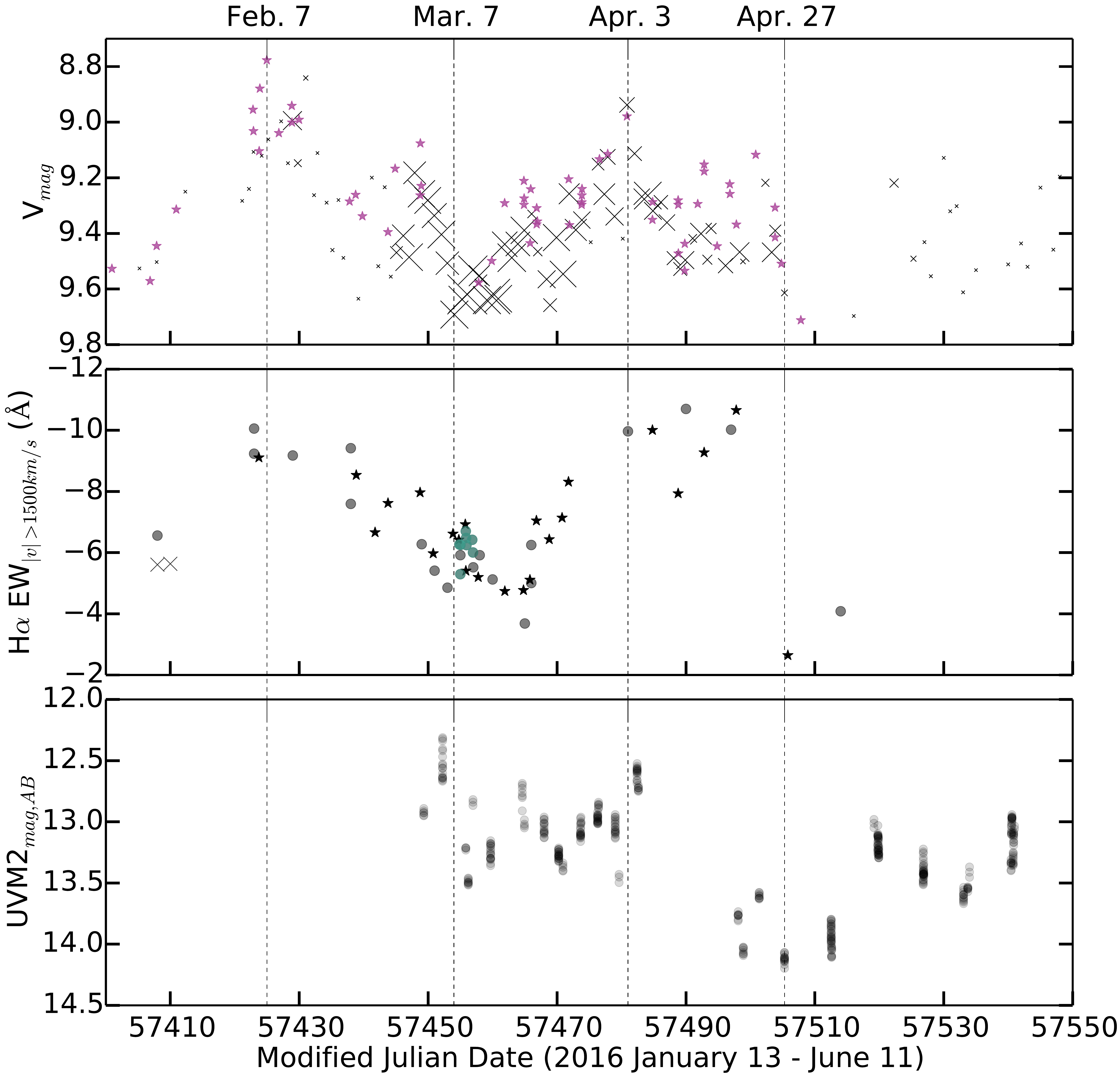}}
\caption{NUV and optical variability during the outflow fast state (2016 January 13 -- April 26) and beyond. The strength of the high velocity optical absorption tracks the optical and NUV flux, consistent with a photoionization effect. The NUV flux varies with the optical flux, sometimes with a larger amplitude.\protect\\ {\bf Top panel}: V band photometry from the AAVSO \citep[grey X points;][]{Kafka2017} and \citet{Munari2016} (purple star points), a crude proxy for accretion rate variability through the visual-emitting disc. The AAVSO data are drawn in 1 day bins (taking the median in linear flux units), and the size of the cross is proportional to the number of contributing observations ranging from 1 to hundreds; the \citet{Munari2016} data are not binned. \protect\\ {\bf 2nd panel}: Equivalent width of the H$\alpha$ absorption trough from blue-shifts faster than -1500 km $^{-1}$,  from R=18000--20000 Echelle spectra (black star points), and R=1300 spectra taken at Asiago and L'Aquilla (grey circle points) and at Liverpool (blue-green circle points). In poorly sampled time periods, data from volunteers published in the ARAS database are also included (grey X points). Spectra were smoothed to a common resolution before measuring the equivalent width. \protect\\ {\bf 3rd panel}: UVM2 Swift event mode photometry for MWC 560 (grey circle points), a crude proxy for accretion rate through the NUV-emitting disc. The data are binned to 60 second intervals. Uncertainties from photon counting statistics are on the scale of the data points, with a median of 0.02 magnitudes. Flickering and systematic errors manifest as streaks; most of these streaks are real flickering, as exemplified in \autoref{UVflick}.} \label{fig3}
\end{figure*}

During the 2016 outflow fast state, a clear correlation was observed between optical/NUV flux and the equivelant width of high-velocity ($|$v$|$\textgreater1500 km~s$^{-1}$) Balmer absorption. This correlation is demonstrated in \autoref{fig3}. The NUV flux varied with the optical flux, sometimes with a larger amplitude. The strength of high-velocity absorption tracked both the optical and NUV flux.

\begin{figure}
\centerline{\includegraphics[width=3.5in]{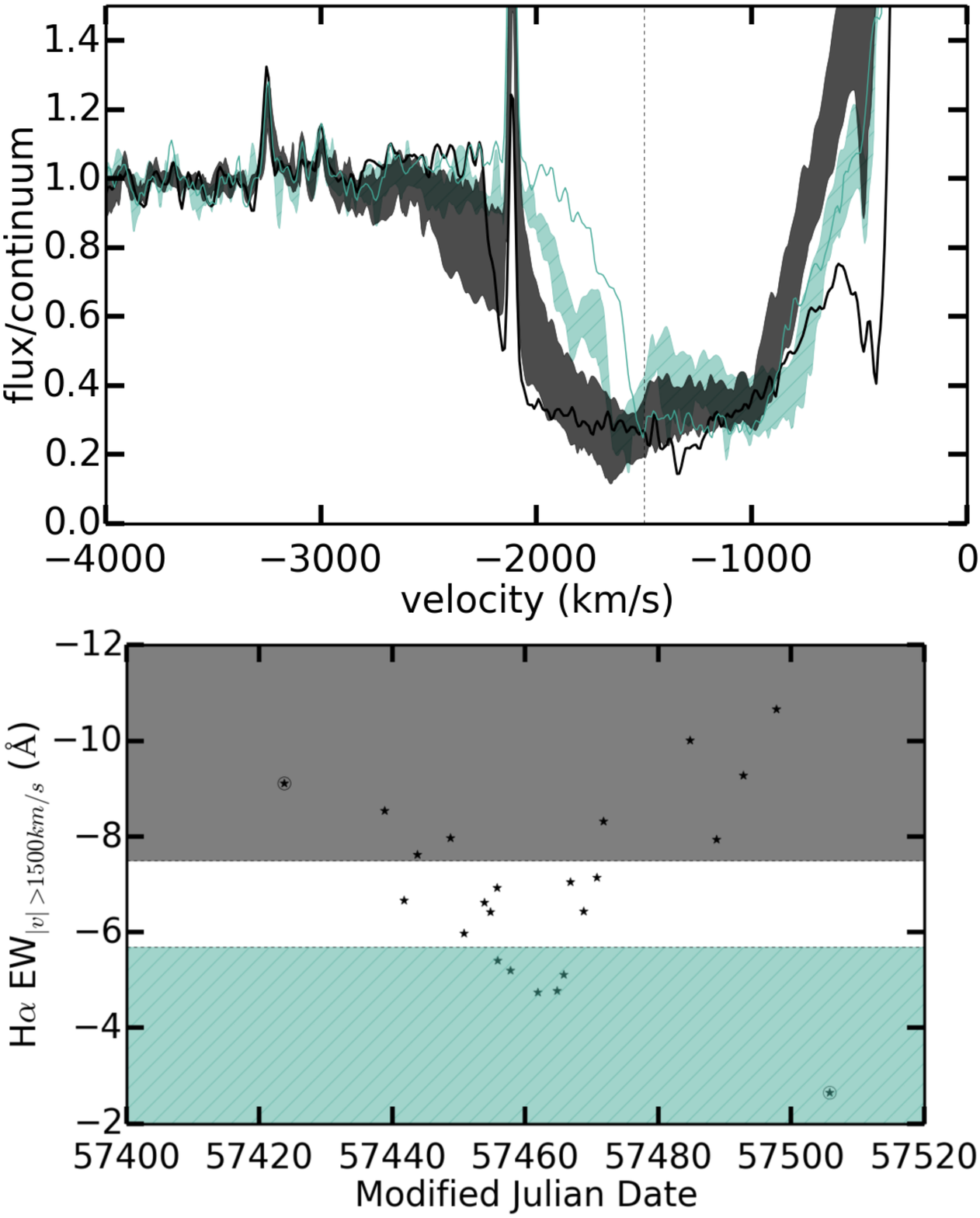}}
\caption{Echelle H$\alpha$ velocity spectra 2016 January through April. The bottom panel mimics the last panel of \autoref{fig3} (including now only echelle spectra) and provides a colour-code for the grey and blue spectra in the top panel. Spectra with H$\alpha$ high-velocity equivalent widths that fall in the bottom panel's blue-shaded small-EW zone are drawn as a shaded-in blue spread in the top panel, and likewise for high-EW zone spectra coded as grey. Two outlier spectra in each zone are marked with enclosing circles in the bottom panel and drawn as blue or black lines in the top panel. Spectra that do not fall into either zone in the bottom panel are not drawn in the top panel.} \label{redgrey}
\end{figure}

\begin{figure}
\centerline{\includegraphics[width=3.5in]{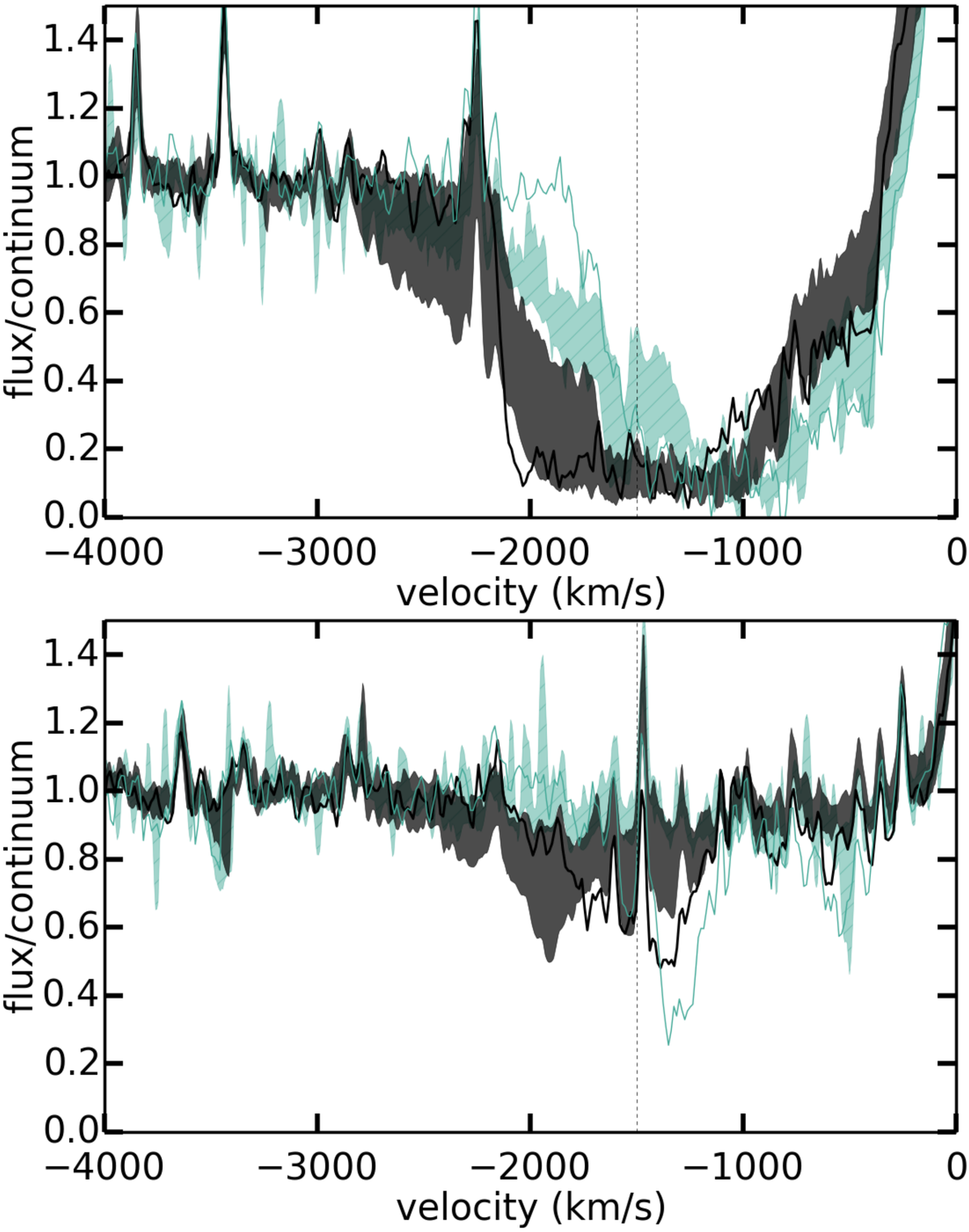}}
\caption{H$\beta$ (top panel) and \ion{Fe}{ii}$\lambda5018$ (bottom panel) velocity spectra, with epochs colour-coded identically to \autoref{redgrey}.} \label{hbeta_feii}
\end{figure}

The correlation between high-velocity Balmer absorption strength and optical/NUV flux is consistent with a photoionization effect. Balmer absorption from n=2 of neutral hydrogen occurs in the partially-ionized zone, just outside the fully-ionized \ion{H}{ii} zone that surrounds the source of ionizing flux. As discussed in \citet{Hamann2019} and \citet{Williams2017}, if the ionizing flux is not high enough to ionize the entire gas slab, then increasing the ionizing flux expands the partially-ionized zone and leads to more Balmer absorption (assuming that the column density is high enough and that the high density requirements to excite to n=2 are met). If the ionizing flux is further increased and the partially-ionized zone reaches the edge of the gas slab, the relationship reverses, and increasing the ionizing flux begins to reduce Balmer absorption and fully-ionize the whole slab. If the ionization parameter in the MWC 560 outflow is relatively low such that the former case applies, then we would expect a correlation between ionizing flux and Balmer absorption strength---which we observe. Assuming that the photoionizing flux was proportional to the optical flux, the amplitude of the optical depth variability is consistent with that expected in much of the parameter space modelled by \citet{Hamann2019} and \citet{Williams2017}.

Alternatively, the correlation could be the result of real variability in the amount of high-velocity mass. If true, this would suggest that the 2016 outflow fast state was a period during which the accretion rate and the outflow power were correlated. But photoionization effects mean we must be cautious in treating the Balmer absorption line as if it were precisely representative of the outflowing mass itself in the absence of other evidence. Fortunately, as discussed in \autoref{4.1.1} and partially thanks to radio and X-ray observations, this is not a significant problem for conclusions reached in the main text of this paper.

In order to detect the correlation between high-velocity Balmer absorption and optical/NUV flux, we smoothed the spectra to a common resolution of R=450. This was necessary due to the wide variety of spectral resolutions employed, the ambiguities of continuum placement, and the blending of emission features into the absorption line. The smoothing was performed using the {\it instrBroadGaussFast} function in the Python AstroLib ({\it pyasl} in PyAstronomy), after first re-binning with flux conserved to a uniform wavelength grid (using\footnote{\url{http://www.astrobetter.com/blog/2013/08/12/python-tip-re-sampling-spectra-with-pysynphot/}~(code~by~J.~Lu)} the {\it spectrum} and {\it observation} modules in {\it pysynphot}) where necessary. The smoothing decreases the absolute value of all the measured equivalent widths by an amount that depends only on the original resolution (i.e., by the same amount for spectra with the same original resolution), up to a few angstroms.

Finally, we present our high-resolution echelle velocity spectra during the 2016 outflow fast state for the main unblended lines for which we have high-SNR coverage: H$\alpha$ (\autoref{redgrey}), and H$\beta$ and \ion{Fe}{ii} (\autoref{hbeta_feii}). The Balmer lines show the same pattern of high-velocity absorption variability seen at low resolution. \ion{Fe}{ii} exhibits a similar trend with more scatter.

\section{No detectable intra-epoch variability in X-rays} \label{app4}

MWC 560 was marginally detected in Swift XRT with less than 3$\sigma$ significance, and we used this to check for variability on time-scales of months; none was detected. We obtained 3$\sigma$ upper limits with the {\it conf} task, after binning to at least 20 source-region counts and subtracting background, to the average Swift spectrum (from 2016 March 2 -- June 1). Freezing all parameters to the Chandra best fit except the normalizations, we obtained upper limits $\approx$2 times the Chandra best fit values, meaning that the average X-ray flux from both components from March 2 through June 1 (sampled at roughly even intervals) was at most double the March 8-9 Chandra best fit. There is no evidence for variability; the Chandra fit was consistent with both the average Swift spectrum and the last month of Swift data (May 5 - June 1).

There was no evidence for X-ray variability within the Chandra observations either, although with only about 5 source counts per kilosecond, the system again was not bright enough for strong constraints. The model fit to the merged spectrum was consistent with both 25 ks March 8 and 9 spectra individually; following the same procedure used to constrain Swift variability, we found that the soft component normalization varied by less than $\pm$50\% from the merged best fit and the hard component normalization varied by less than +100\% and -75\% from the merged best fit, between March 8 and 9 at 3$\sigma$. Likewise, visual inspection of full-energy-band Chandra light curves obtained with {\it dmextract} for a variety of time bins did not reveal any instances of variability in excess of the expected statistical noise. For that inspection, we estimated 1$\sigma$ errors using the \citet{Gehrels1986} Equation 7 upper margin with S=1, a standard option in CIAO; the average statistical uncertainty for a single bin was 80\%, 25\%, and 15\% of the source count rate for 1, 5, and 12.5 ks bins, respectively. Finally, for each binning, we calculated the observed standard deviation of the whole light curve and found it to be consistently 45-60\% of these statistical uncertainties.

\section{X-ray fitting routine details} \label{app5}

We used {\it chi2gehrels} statistics \citep[low-count Poisson statistics from][]{Gehrels1986} and the {\it moncar} method on the background-subtracted spectrum. The {\it moncar} method (a Monte Carlo routine that draws fit statistics for one variable at a time from parameter space, while allowing all other variables to float to their best fit at that point in parameter space) was motivated by a non-parabolic distribution of fit solutions in parameter space. To accommodate the extended, mildly bi-modal spread of allowable solutions, we increased the calculated confidence intervals to 95$\%$ ($\sim$2$\sigma$) confidence intervals, bracketing the solution spread by requiring a $\chi^2$ at least 3.84 above the minimum. The reduced $\chi^{2}_{\nu}$ is a little high (1.6 at the best fit), but visual inspection of \autoref{xrayspectra} suggests that the fit suits the data and further model components are not motivated. 
 
We checked our analysis against another common data reduction method, obtaining consistent results\footnote{The best fits remained identical to within 1\% for the soft component and 15\% for the hard component, well within the uncertainties. The only more substantial change was a 100\% and 25\% stretching of the upper bound margins for the normalizations and for the hard component absorption, respectively, which does not impact our conclusions.} when using a spectrum extracted from a larger 2.5 arcsec radius region and fitting from 0.3 to 10.0 keV.

The results for variability in observed flux between epochs, included in the main text, were obtained as follows. We froze all parameters to the 2016 best fit except for the normalization of the soft component, which was allowed to vary freely. When this frozen 2016 best fit was applied to the 2013 spectrum, the soft component normalization had a $1\sigma$ minimum of 0 and maximum of $2.5\times10^{-6}$ (best fit $1.5\times10^{-6}$). This range includes results from binning to both 20- and 40- source-region counts per bin (the latter avoids bins with negative counts in the background-subtracted spectrum, and gives our quoted best fit value).  Trying to fit the background introduced too many free parameters to be useful, but we did try a version of low-count Cash statistics ({\it wstat} in {\it Sherpa}) on the un-binned background-subtracted spectrum and obtained a much smaller maximum. Similarly, when the frozen 2016 best fit was applied to the 2016 spectrum with Gehrels $\chi^2$ statistics, the soft component normalization was $1.9_{-0.2}^{+0.3}\times10^{-5}$, almost independent of binning. In this way, we obtained that between the 2013 and 2016 X-ray epochs, the observed soft component flux increased by a factor of 13 (1$\sigma$ lower bound of 7).

For the inter-epoch variability in the hard component, an essentially identical method was employed: freezing the soft component to the 2013 fit obtained in the last section, freezing all hard component parameters except the normalization to the 2007 \citet{Stute2009} fit, and additionally fixing the weak Gaussian emission line component included in the \citet{Stute2009} model to have a normalization proportional to the hard component normalization.

Throughout the main text, observed X-ray flux and its derived quantities are quoted with rough 1$\sigma$ uncertainties from the fits with only 1 free parameter, while X-ray temperature and absorption columns are quoted with rough 2$\sigma$ uncertainties from the fits with 5 free parameters.

\section{Optical loading constraints} \label{app6}

X-ray telescopes are susceptible to optical loading, in which optical photons reach the detector and masquerade as X-ray photons; red/infrared leaks can let optical light through the shielding. Symbiotic stars are particularly dangerous; not only do they contain bright cool giants with high red/infrared flux, they also contain colliding winds that produce soft X-rays with about the same energies as optical loading typically produces. Fortunately, the leaks and detectors on Swift and Chandra are well-enough characterized that we can check whether optical loading is an issue for any given object; here, we do so for MWC 560.

We ruled out optical loading in our Chandra observations. The Chandra optical leak is between the I and J bands. On the S3 chip, for sources dimmer than 5 magnitudes in those bands, no more than $\sim1$ contaminant ADU appears on the brightest pixel per frame, much less than the $\approx20$ ADU pixel$^{-1}$ required to be interpreted as an event count and much less than the 13 ADU pixel$^{-1}$ split threshold used for event grading; also, the 0.004 keV energy of an ADU is not enough to change the energy of real events significantly\footnote{\url{http://cxc.harvard.edu/cal/Hrma/UvIrPSF.html}}. The J band magnitude of MWC 560 is about 6.5 \citep{Meier1996}, and I band magnitudes around 8 were recorded by the AAVSO during the 2016 outflow fast state \citep{Kafka2017}.

It is interesting to note that optical loading on Chandra could be a very important issue for those symbiotic stars with I and J magnitudes brighter than -0.4, especially any observations taken on the S3 chip with a standard frame length early on in the life of the Chandra mission. As of 2006, that magnitude corresponded to the split threshold. Over time, Chandra's filter has developed a thickening layer of contaminants that has helped plug the red leak; there is some evidence that it now takes a much brighter source to reach the split threshold, though these new tests have not yet been formally characterized (Chandra help desk 2017, private communication).

We estimated the optical loading in Swift XRT using the online tool provided by Swift\footnote{\url{http://www.swift.ac.uk/analysis/xrt/optical_tool.php}}, which required us to choose a T$_{eff}$ and M$_{bol}$-M$_{v}$ for the source, and we adopted T$_{eff}$=3330 K and M$_{bol}$-M$_{v}$=-2.15. A PC-mode XRT count threshold of 10$^{-4}$ counts s$^{-1}$ yield a V band threshold of 9.7 magnitudes, so the giant is dim enough not to cause optical loading on its own. The outflow fast state accretion disc is typically much bluer, and probably not an issue (the same threshold for a B0 star is V=8.6). However, the fast state accretion disc colour is not especially well characterized, particularly longer than $8000$\AA. So we cannot totally rule out optical loading in the Swift XRT, especially during the brightest phases of the 2016 high state. Fortunately, our uncertainty on this point is not of great concern, because we only used the Swift XRT to check for variability (of which we found none).

\section{Estimating the accretion luminosity} \label{app7}

\begin{figure}
\centerline{\includegraphics[width=3.5in]{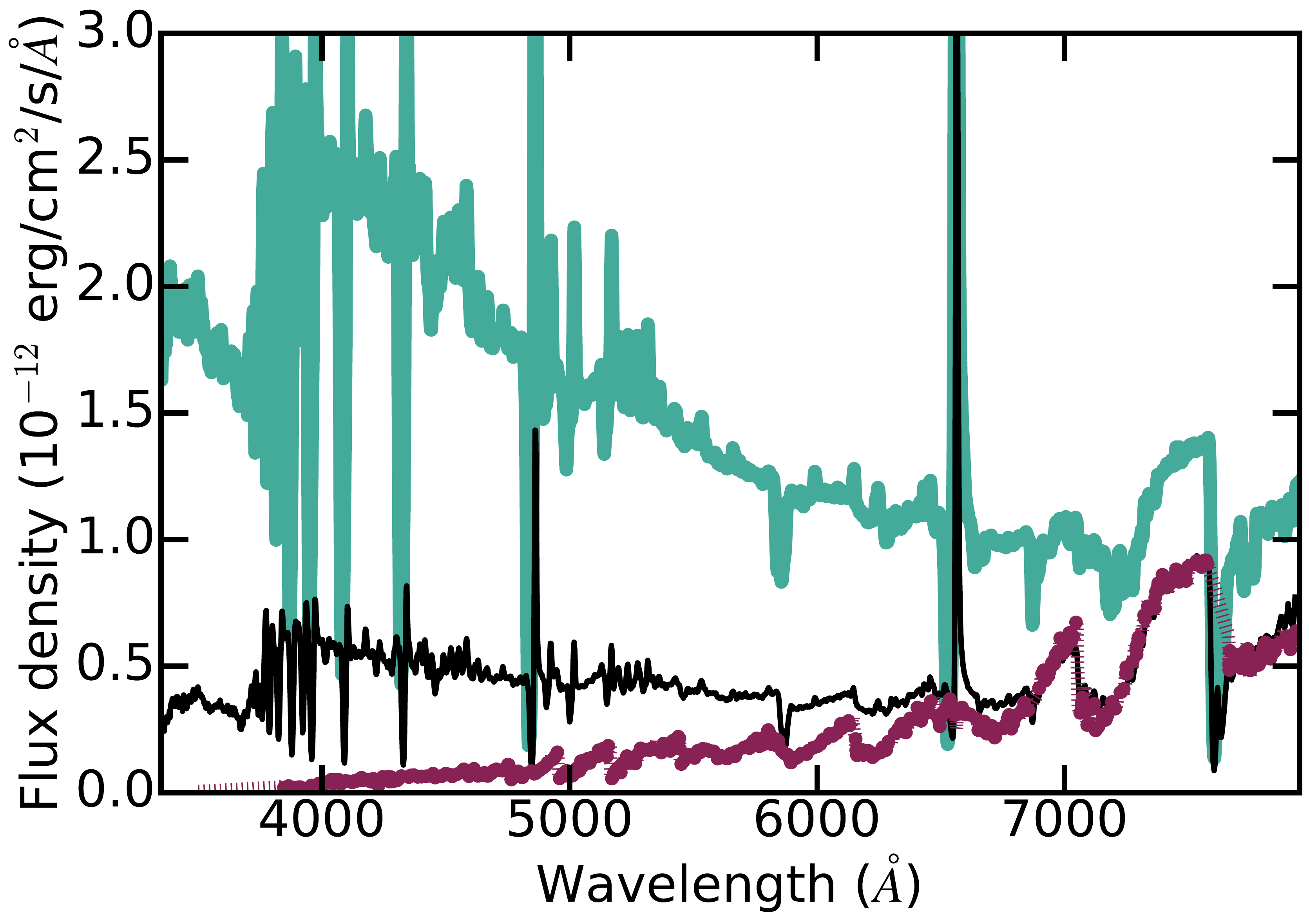}}
\caption{Optical spectra of MWC 560 used to calculate luminosities: the brightest outflow fast state spectrum (2016 Apr 3), a quiescent state spectrum (2015 Mar 10) typical for the present decade, and the Case M5 giant template from \citet{Fluks1994} scaled by eye to match the red side of the quiescent spectrum (which leads to the smoothest residuals). The MWC 560 spectra are corrected for extinction with E(B-V)$=0.15$ (Appendix~\ref{app1}). The differing spectral slopes are primarily due to the additive contribution of the giant, which has a smaller fractional contribution during the outflow fast state. Only low-resolution spectra are plotted here, decreasing the prominence of narrow metal emission lines.} \label{SED}
\end{figure}

To estimate the accretion disc luminosity at the 2016 peak and during quiescence, we took our spectra from 2016 April 3 (the brightest spectrum) and 2015 March 10, and dereddened them by E(B-V)$=0.15$ (Appendix~\ref{app1}) with a \citet{Fitzpatrick2007} reddening model using K. Barbary's {\it extinction} python code\footnote{\url{https://extinction.readthedocs.io/en/latest/index.html}}. We then took the Case M5 RG template \citep{Fluks1994} and scaled it by eye to match the red side of the {\it quiescent} spectrum (which leads to the smoothest residuals in both spectra), then subtracted this constant RG spectrum from the observed spectra before integrating the optical luminosity. The pre-subtraction dereddened spectra and the scaled RG template are shown in \autoref{SED}.

Our best guess accretion rate for the 2016 peak is 6 $\times 10^{-7}$ (d/2.5 kpc)$^{2}$ (R / 0.01 R$_{\odot}$) (0.9 M$_{\odot}$/ M) M$_{\odot}$ yr$^{-1}$, obtained as follows. The peak optical (3300--7930\AA) luminosity in the 2016 outflow fast state was about 1100 L$_{\odot}$ (d/2.5 kpc)$^2$. The peak Swift NUV flux (2100--2870\AA) was about 70 L$_{\odot}$ (d/2.5 kpc)$^2$, but judging by the shape of the NUV spectrum (see \autoref{uvspectrasec}), we suspect that the intrinsic unabsorbed 1200--3300\AA\ range had a luminosity similar to the April 29 NUV+FUV IUE spectra, or about 730 (d/2.5 kpc)$^2$ L$_{\odot}$.  Our total peak accretion disc luminosity is therefore about the same as that of the 1990 outburst \citep{Tomov1992,Schmid2001}.

We estimate a quiescent accretion rate minimum value of about 1 $\times$ 10$^{-7}$(d/2.5 kpc)$^2$ (R / 0.01 R$_{\odot}$) (0.9 M$_{\odot}$/ M) M$_{\odot}$ yr$^{-1}$. This comes from about 205 (d/2.5 kpc)$^2$ L$_{\odot}$ observed from 3300--7930\AA\ in our RG-subtracted optical spectrum from 2015 March 8, 65 (d/2.5 kpc)$^2$ L$_{\odot}$ observed from 1200--3300\AA\ in the 1993 February 23 SWP and LWP spectra observed by the IUE during a quiescent period with similar optical flux, and an estimated 30 (d/2.5 kpc)$^2$ L$_{\odot}$ absorbed by the \ion{Fe}{ii} curtain. We ignore the possibility that up to half the luminosity could be emitted by the boundary layer in the extreme-ultraviolet and not re-emitted, which if true would double our estimated accretion rates.


\bsp	
\label{lastpage}
\end{document}